\documentclass[twocolumn]{aastex63}
\usepackage{natbib}
\usepackage{graphicx}
\usepackage{tikz}
\usepackage{enumitem}
\usepackage{float}
\usepackage{textcomp}
\usepackage{makecell}
\usepackage{comment}
\usepackage{amsmath} 

\usepackage{totcount}
\graphicspath{{./}{}}
\usepackage{fontspec}
\usepackage[english]{babel}
\babelprovide{punjabi}
\babelfont[punjabi]{rm}[Renderer=Harfbuzz,Script=Gurmukhi]{FreeSerif.otf}
\babelprovide{chinese-simplified}
\babelfont[chinese-simplified]{rm}[Renderer=Harfbuzz,Script=CJK,Scale=MatchLowercase,Path=./]{NotoSerifCJKsc-sub.otf}
\newcommand{\Punj}[1]{\foreignlanguage{punjabi}{#1}}
\newcommand{\CJKtext}[1]{\foreignlanguage{chinese-simplified}{#1}}

\newcommand{\code}[1]{\texttt{#1}}

\newcommand{\JWST}{\textit{JWST}~}
\newcommand{\jwst}{\textit{JWST}~}
\newcommand{\HST}{\textit{HST}~}

\newcommand{\logM}{log($M_\star/M_\odot$) \,}

\newcommand{\ewha}{EW($\mathrm{H\alpha + N[II]}$)}

\newcommand{\logm}{log$_{10}$(M$_*$/M$_\odot$)}

\newtotcounter{citnum} 
\def\oldbibitem{} \let\oldbibitem=\bibitem
\def\bibitem{\stepcounter{citnum}\oldbibitem}

\newcommand{\loosera}{\cite{Looser2023a}~}
\newcommand{\looserb}{\cite{Looser2023b}~}
\newcommand{\strait}{\cite{Strait2023}~}

\date{\today}

\submitjournal{ApJ}

\shorttitle{UNCOVER Napping Galaxies at $z\sim4-7$}
\shortauthors{Khullar et al. 2026}

\begin{document}

\title{Caught Napping by \textit{JWST} UNCOVER/MegaScience: Constraining bursty star formation histories and number densities of mini-quenched galaxies at redshifts 4-7}

\author[0000-0002-3475-7648]{Gourav Khullar (\Punj{ਗੌਰਵ ਖੁੱਲਰ})}
\altaffiliation{Baum Postdoctoral Fellow for Innovative Astronomy}
\affiliation{Department of Astronomy, University of Washington, Physics-Astronomy Building, Box 351580, Seattle, WA 98195-1700, USA}
\affiliation{eScience Institute, University of Washington, Physics-Astronomy Building, Box 351580, Seattle, WA 98195-1700, USA}
\affiliation{Department of Physics and Astronomy and PITT PACC, University of Pittsburgh, Pittsburgh, PA 15260, USA}

\author[0000-0001-5063-8254]{Rachel Bezanson}
\affiliation{Department of Physics and Astronomy and PITT PACC, University of Pittsburgh, Pittsburgh, PA 15260, USA}

\author[0000-0002-1714-1905]{Katherine A. Suess}
\affiliation{Department for Astrophysical \& Planetary Science, University of Colorado, Boulder, CO 80309, USA}
\affiliation{Department of Astronomy and Astrophysics, University of California, Santa Cruz, 1156 High Street, Santa Cruz, CA 95064 USA}

\author[0000-0001-7300-9450]{Ikki Mitsuhashi}
\affiliation{Department for Astrophysical \& Planetary Science, University of Colorado, Boulder, CO 80309, USA}

\author[0000-0003-4075-7393]{David J. Setton}\thanks{Brinson Prize Fellow}
\affiliation{Department of Astrophysical Sciences, Princeton University, Princeton, NJ 08544, USA}

\author[0000-0001-6755-1315]{Joel Leja}
\affiliation{Department of Astronomy \& Astrophysics, The Pennsylvania State University, University Park, PA 16802, USA}
\affiliation{Institute for Computational \& Data Sciences, The Pennsylvania State University, University Park, PA 16802, USA}
\affiliation{Institute for Gravitation and the Cosmos, The Pennsylvania State University, University Park, PA 16802, USA}

\author[0000-0002-0108-4176]{Sedona H. Price}
\affiliation{Space Telescope Science Institute, 3700 San Martin Drive, Baltimore, MD 21218, USA}
\affiliation{Department of Physics and Astronomy and PITT PACC, University of Pittsburgh, Pittsburgh, PA 15260, USA}

\author[0000-0001-7160-3632]{Katherine E. Whitaker}
\affiliation{Department of Astronomy, University of Massachusetts, Amherst, MA 01003, USA}
\affiliation{Cosmic Dawn Center (DAWN), Niels Bohr Institute, University of Copenhagen, Jagtvej 128, K{\o}benhavn N, DK-2200, Denmark} 

\author[0000-0001-8174-317X]{Emilie Burnham}
\affiliation{Department of Astronomy \& Astrophysics, The Pennsylvania State University, University Park, PA 16802, USA}
\affiliation{Institute for Gravitation and the Cosmos, The Pennsylvania State University, University Park, PA 16802, USA}
\affiliation{Center for Astrostatistics and Astroinformatics, The Pennsylvania State University, University Park, PA 16802, USA}

\author[0000-0003-1614-196X]{John R. Weaver}\thanks{Brinson Prize Fellow}
\affiliation{MIT Kavli Institute for Astrophysics and Space Research, 70 Vassar Street, Cambridge, MA 02139, USA}

\author[0009-0009-9795-6167]{Iryna Chemerynska}
\affiliation{Institut d'Astrophysique de Paris, UMR 7095, CNRS, and Sorbonne Universit\'e, 98 bis boulevard Arago, 75014 Paris, France}

\author[0000-0001-6278-032X]{Lukas J. Furtak}
\affiliation{Cosmic Frontier Center, The University of Texas at Austin, Austin, TX 78712, USA}
\affiliation{Department of Astronomy, University of Texas at Austin, 2515 Speedway, Austin, Texas 78712, USA}

\author[0000-0002-5612-3427]{Jenny Greene}
\affiliation{Department of Astrophysical Sciences, Princeton University, Princeton, NJ 08544, USA}

\author[0000-0001-9269-5046]{Bingjie Wang (\CJKtext{王冰洁})}
\thanks{NHFP Hubble Fellow}
\affiliation{Department of Astrophysical Sciences, Princeton University, Princeton, NJ 08544, USA}

\author[0000-0002-7570-0824,sname=Atek,gname=Hakim]{Hakim Atek}
\affiliation{Institut d'Astrophysique de Paris, UMR 7095, CNRS, and Sorbonne Universit\'e, 98 bis boulevard Arago, 75014 Paris, France}

\author[0000-0003-2680-005X]{Gabe Brammer}
\affiliation{Cosmic Dawn Center (DAWN), Niels Bohr Institute, University of Copenhagen, Jagtvej 128, K{\o}benhavn N, DK-2200, Denmark}

\author[0000-0003-3881-1397]{Olivia R. Cooper}\altaffiliation{NSF Astronomy and Astrophysics Postdoctoral Fellow}
\affiliation{Department for Astrophysical \& Planetary Science, University of Colorado, Boulder, CO 80309, USA}

\author[0000-0002-1109-1919]{Robert Feldmann}
\affiliation{Department of Astrophysics, University of Zurich, CH-8057, Switzerland}

\author[0000-0001-7201-5066]{Seiji Fujimoto}
\affiliation{David A. Dunlap Department of Astronomy and Astrophysics, University of Toronto, 50 St. George Street, Toronto, Ontario, M5S 3H4, Canada}
\affiliation{Dunlap Institute for Astronomy and Astrophysics, 50 St. George Street, Toronto, Ontario, M5S 3H4, Canada}

\author[0000-0002-2380-9801]{Anna de Graaff}
\affiliation{Max-Planck-Institut f\"ur Astronomie, K\"onigstuhl 17, D-69117 Heidelberg, Germany}
\affiliation{Center for Astrophysics $|$ Harvard \& Smithsonian, 60 Garden St., Cambridge MA 02138 USA}\thanks{Clay Fellow}

\author[0000-0002-2057-5376]{Ivo Labbe}
\affiliation{Centre for Astrophysics and Supercomputing, Swinburne University of Technology, Melbourne, VIC 3122, Australia}

\author[0000-0001-9002-3502]{Danilo Marchesini}
\affiliation{Department of Physics and Astronomy, Tufts University, 574 Boston Ave., Medford, MA 02155, USA}

\author[0000-0002-2446-8770]{Ian McConachie}
\affiliation{Department of Astronomy, University of Wisconsin-Madison, 475 N. Charter St., Madison, WI 53706 USA}

\author[0000-0001-8367-6265]{Tim B. Miller}
\affiliation{Center for Interdisciplinary Exploration and Research in Astrophysics (CIERA), Evanston, IL 60201, USA}

\author[0000-0002-9816-9300]{Abby Mintz}
\affil{Department of Astrophysical Sciences, Princeton University, 4 Ivy Lane, Princeton, NJ 08544, USA}

\author[0000-0003-2804-0648]{Themiya Nanayakkara}
\affiliation{Sydney Institute for Astronomy, School of Physics A28, The University of Sydney, NSW 2006, Australia.}

\author[0000-0001-5851-6649]{Pascal Oesch}
\affiliation{Department of Astronomy, University of Geneva, Chemin Pegasi 51, 1290 Versoix, Switzerland}
\affiliation{Cosmic Dawn Center (DAWN), Niels Bohr Institute, University of Copenhagen, Jagtvej 128, K{\o}benhavn N, DK-2200, Denmark}

\author[0000-0002-9651-5716]{Richard Pan}\affiliation{Department of Physics and Astronomy, Tufts University, 574 Boston Ave., Medford, MA 02155, USA}

\author[0000-0001-7503-8482]{Casey Papovich}
\affiliation{Department of Physics and Astronomy, Texas A\&M University, College Station, TX, 77843-4242 USA}

\author[0000-0001-6454-1699]{Yunchong Zhang}
\affiliation{Department of Physics and Astronomy and PITT PACC, University of Pittsburgh, Pittsburgh, PA 15260, USA}

\correspondingauthor{Gourav Khullar}
\email{gkhullar@uw.edu}

\begin{abstract}

We explore the prevalence of mini-quenched or ``napping'' galaxies selected from spectroscopic and photometric samples in the UNCOVER/MegaScience survey. These galaxies are empirically identified by the presence of moderate Balmer breaks, weak emission lines ($EW(H\alpha) < 100$\AA) and relatively blue UV continua. We infer the star formation histories (SFHs) of our sample using flexible non-parametric models with \code{Prospector} optimized to capture recent episodes of bursty star formation and quenching, and find that they are uniquely identifiable in the SFR$_{10}$/SFR$_{100}$ parameter space moving towards (temporary) quiescence. We demonstrate that although spectroscopy is best able to identify rapidly declining SFRs, densely sampled medium-band photometry recover these key spectral features and thus robustly identify pure samples of this transient phase -- with imaging alone. We quantify the number density of napping galaxies at $z=4-7$ in the Abell 2744 lensing field, finding 8 spectroscopically confirmed nappers and 60 photometric candidates spanning \logM $= 7.5-10$. We verify that the photometry alone can identify a pure sample of nappers, leveraging a smaller high signal-to-noise ratio spectroscopic sample. Consistent with previous studies, we find that nappers are most common at low stellar mass (\logM$\sim9$). We see a hint that the number densities increase from $z\sim6$ to $z\sim4$, though our small sample is likely affected by cosmic variance. Our study demonstrates the increasing importance of stochastic star formation as a regulator of low-mass galaxy growth in the several hundred Myr after reionization, and offers a direct observational testbed for the strength and duty cycle of stellar feedback in cosmological simulations.

\end{abstract}

\keywords{quiescent galaxies, gravitational lensing, \JWST, star formation histories, SED fitting, burstiness, post-starburst galaxies}

\section{Introduction}
\label{sec:intro}

At high redshifts --- $z>4$, when the Universe was $\sim$ 1 Gyr old or younger --- a galaxy's star formation history (SFH) is expected to transition from ``stochastic'' (or bursty) to ``secular'' (e.g., \citealt{Dayal2013,faisst2019,caplar2019, ciesla2024, dome2024, ciesla2024}) especially in the low stellar-mass regime, leading to lower redshift ($3<z<4$) observations of a relatively tighter star forming main sequence (SFMS)\citep{Brinchmann2004,Speagle2014, Schreiber2015,Renzini2015,Leja2020,Leja2022}. Simulations expect this transition to be primarily caused by the deepening of gravitational potentials such that they withstand feedback-based ejection of star-forming gas \citep{Dayal2013,faisst2019,Wilkins2023,Hopkins2023}. This is especially seen in lower stellar mass (\logm$< 9$) galaxies \citep{Dayal2013,ciesla2024,munoz2026}. While \citet{ciesla2024} find that the transition from stochastic to secular star formation occurs already by $z\sim9$ for massive galaxies (log(M$_{\star}$/M$_{\odot}$)$\geq$9), their lower-mass bins (log(M$_{\star}$/M$_{\odot}$)$\geq$8.6) remain stochastic-dominated even at their lowest probed redshift ($z\sim6-7$). This raises the question of whether -- and at what redshift-- a similar transition eventually occurs for lower-mass galaxies.

Therefore, at $z>4$, we expect to observe galaxies with bursty star formation histories, including in "quenched" states, i.e., where the galaxy happens to be in a state of low star formation (SF) in its cycle of fluctuation around the SFMS. Our ability to constrain this mode of star formation and mass assembly is a function of the wavelength coverage and spectral resolution of the observations, survey depth that can capture the ``lull" phase of bursty SFHs, and modeling techniques that constrain and quantify burstiness in a galaxy's SFH \citep{wang2025, burnham2026}. 

JWST has completely opened up the parameter space of measurements/observations of these stochastic SFHs, including ``mini-quenched"/napping galaxies \citep{Looser2023a, Looser2023b,cole2023,whitler:22,trussler2024,Endsley2025,covelo2025}, with multiple spectroscopically confirmed discoveries. These nappers -- most likely O- and B-type post-starburst galaxies (PSBs) -- should contain residual or low amounts of SF, as well as old stellar populations; in galaxy spectra, these features are characterized by bright rest-frame UV flux, a weak Balmer break, and weak or absent emission lines (e.g., \ewha). However, despite the success of these serendipitous discoveries, much remains to be learned about the details of these systems. Nappers are relatively rare but can also be hard to identify robustly \citep{trussler2024} -- however, improvements on their number densities would help constrain the timescale and amplitude of their star formation \citep{mintz2025}. Secondly, generalizing the physical mechanisms involved in causing this napping phase in galaxies is difficult -- supernova feedback acts on $>20-30$ Myr timescales (e.g., \citealt{faucher2018}), and is not necessarily able to explain instances of abrupt quenching \citep{faisst2019,lovell2021,ciesla2024,trussler2024,dome2024,munoz2026}; merger-driven starbursts and other interactions in overdense fields are strong but unconfirmed candidates. Finally, building a sample of such systems photometrically is a challenge -- existing color-color diagnostics do not constrain the combination of weak/moderate Balmer breaks with weak emission lines \citep{Antwi-Danso2022,trussler2024}, requiring unique cuts/rest-frame synthetic filters to constrain the combination of spectral features seen in napping galaxies. 

In order to better understand the impact of stochastic growth histories and refine formation models, we must: a) unpack the properties of an increasingly diversifying population of mini-quenched/napping galaxies observed at $z>4$, and b) build a systematically selected sample of mini-quenched systems at high-redshifts via \JWST-based mass-selected spectroscopic and photometric surveys alike. 

JWST deep photometric and spectroscopic surveys are well suited for the task of solving the above problems. The wavelength coverage and sensitivity of NIRSpec/PRISM mode \citep{nirspec2022}, the depths achieved with NIRCam imaging, and boosts from strong gravitational lensing through the UNCOVER -- Ultradeep NIRSpec and NIRCam ObserVations before the Epoch of Reionization-- Treasury Survey (PIs: Bezanson and Labbe; \citealt{bezanson2022,Suess2024}) is designed to extend the search for quiescent galaxies behind the massive galaxy cluster Abell 2744 (A2744) to unprecedented low masses and high redshifts, i.e., ${<} 10^9M_{\sun}$ to $z{=}9$, including nappers.

In this study, we characterize via spectrophotometric spectral energy distribution (SED) modeling the bursty star formation histories of so-called mini-quenched or napping \footnote{ We use the term ``mini-quenched", ``napping" galaxies, ``nappers" to mean the same kind of galaxy, to signify systems that have low SF at the epoch of observation, but do not show evidence of permanent quenching, as is seen at lower redshifts ($z<2$) and in relatively higher mass (logM $>$10.5) galaxies; from hereon, we use the term "nappers" throughout the paper.} (B-type post-starburst) galaxies at $z=4-7$, first detected in the spectroscopic survey component of UNCOVER. This sample widens the mass and redshift range where these galaxies are observed, and doubles the number of spectroscopically confirmed ``napping'' systems close to the epoch of reionization. We also lay out a prescription for discovering such systems in photometric surveys, and demonstrate that using traditional SFH modeling methodologies with purely photometric observations in wide surveys is not reliable at finding a pure sample of nappers. We use our novel methodology to showcase sample statistics of medium-band photometry-detected nappers as observed in the UNCOVER/MegaScience survey.

\begin{figure*}
\centering
\begin{minipage}[t]{0.48\textwidth}
    \centering
    \includegraphics[width=\linewidth]
    {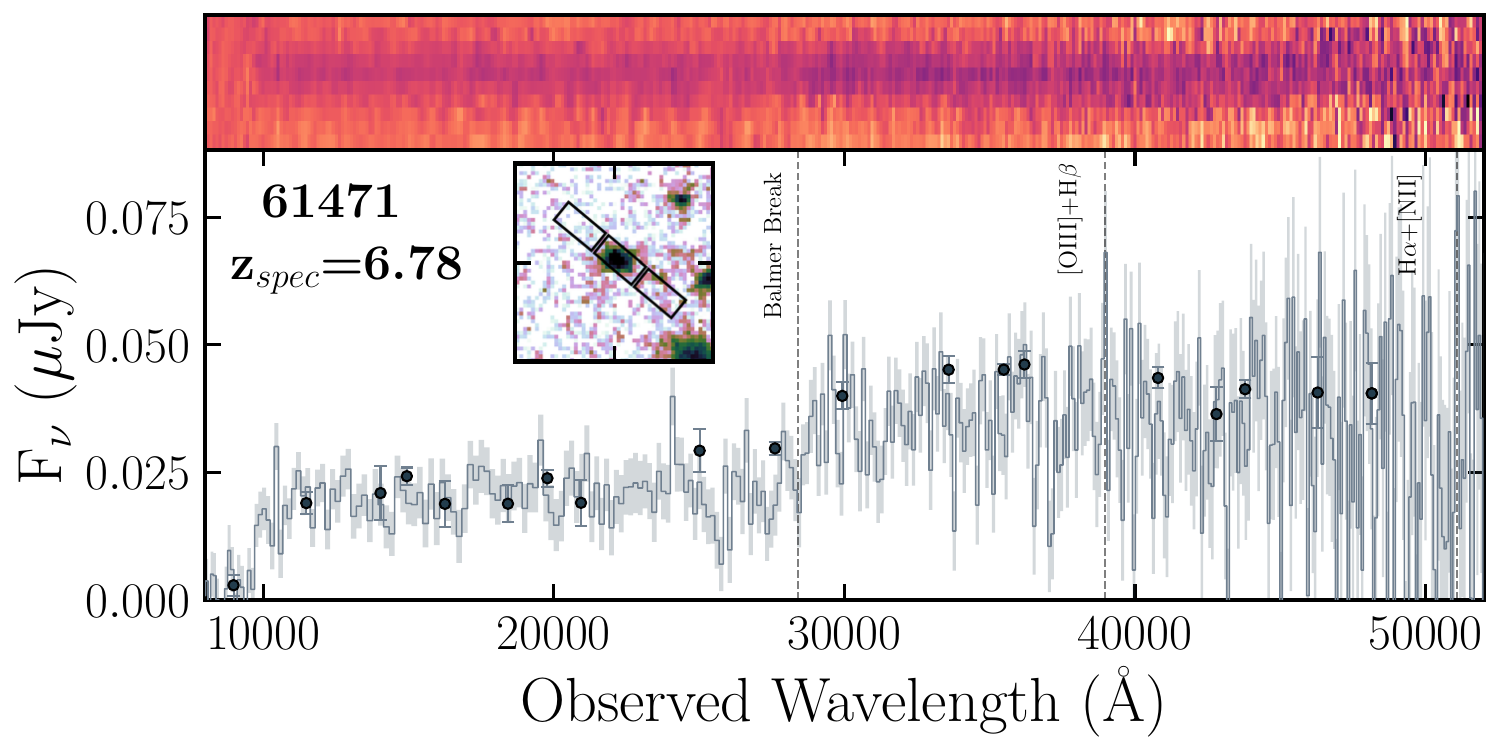}
    \vskip-5mm  
    \includegraphics[width=\linewidth]{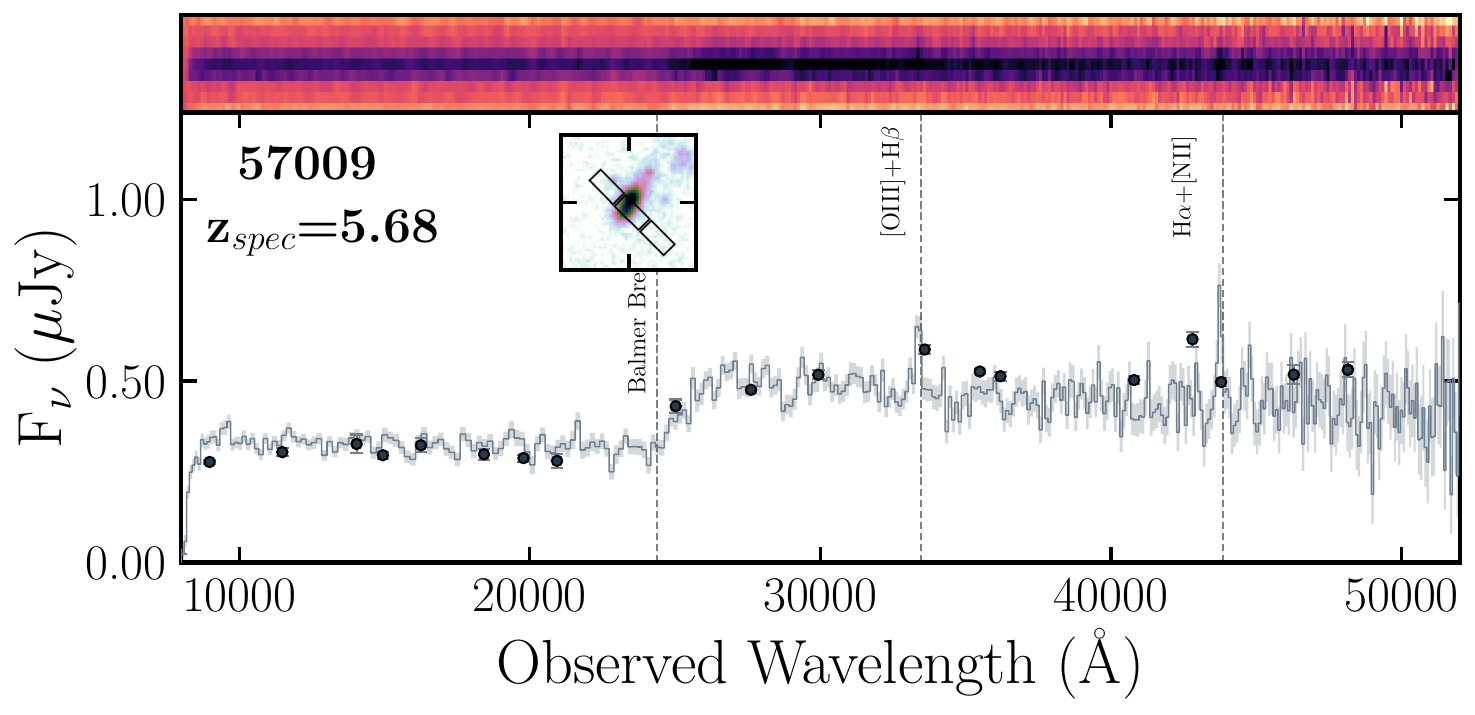}
    \vskip-5mm 
    \includegraphics[width=\linewidth]{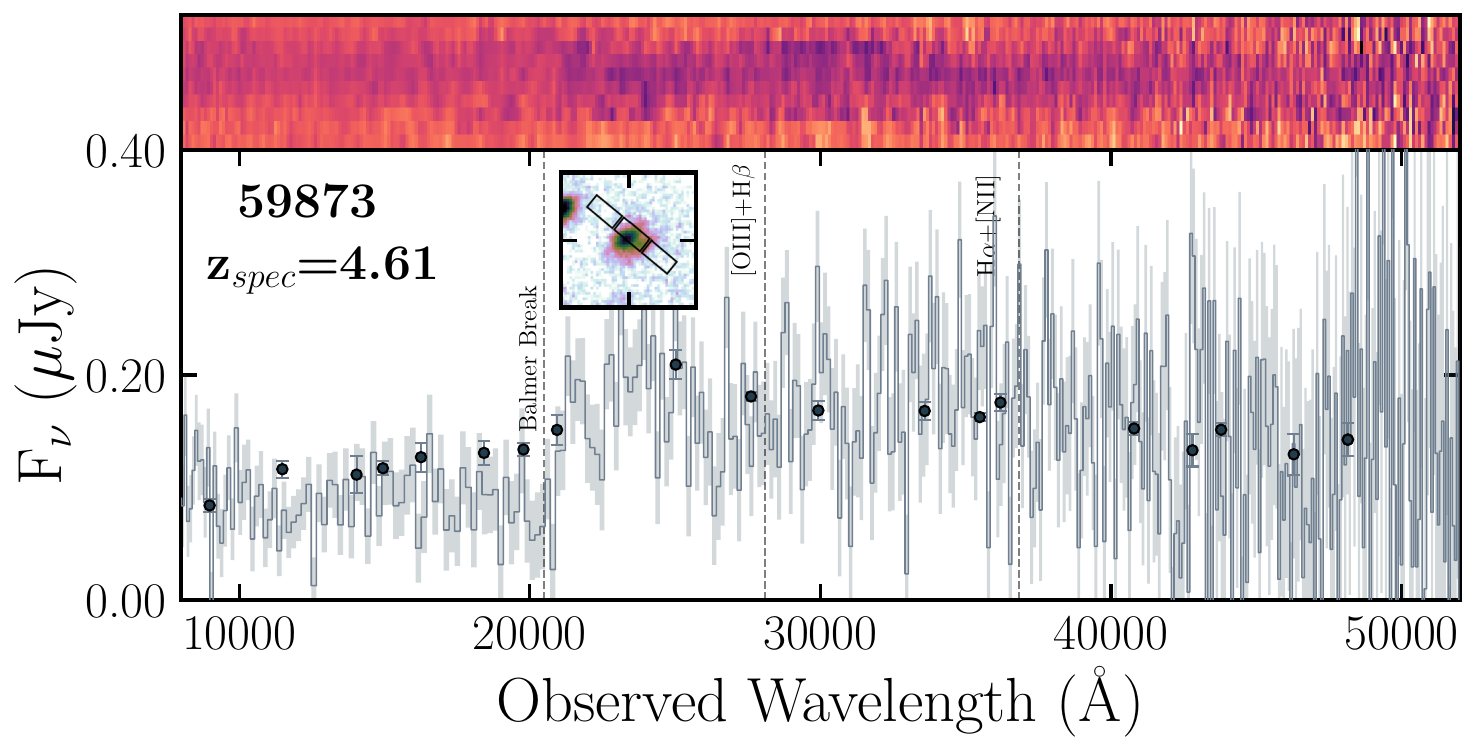}
    \vskip-5mm
    \includegraphics[width=\linewidth]{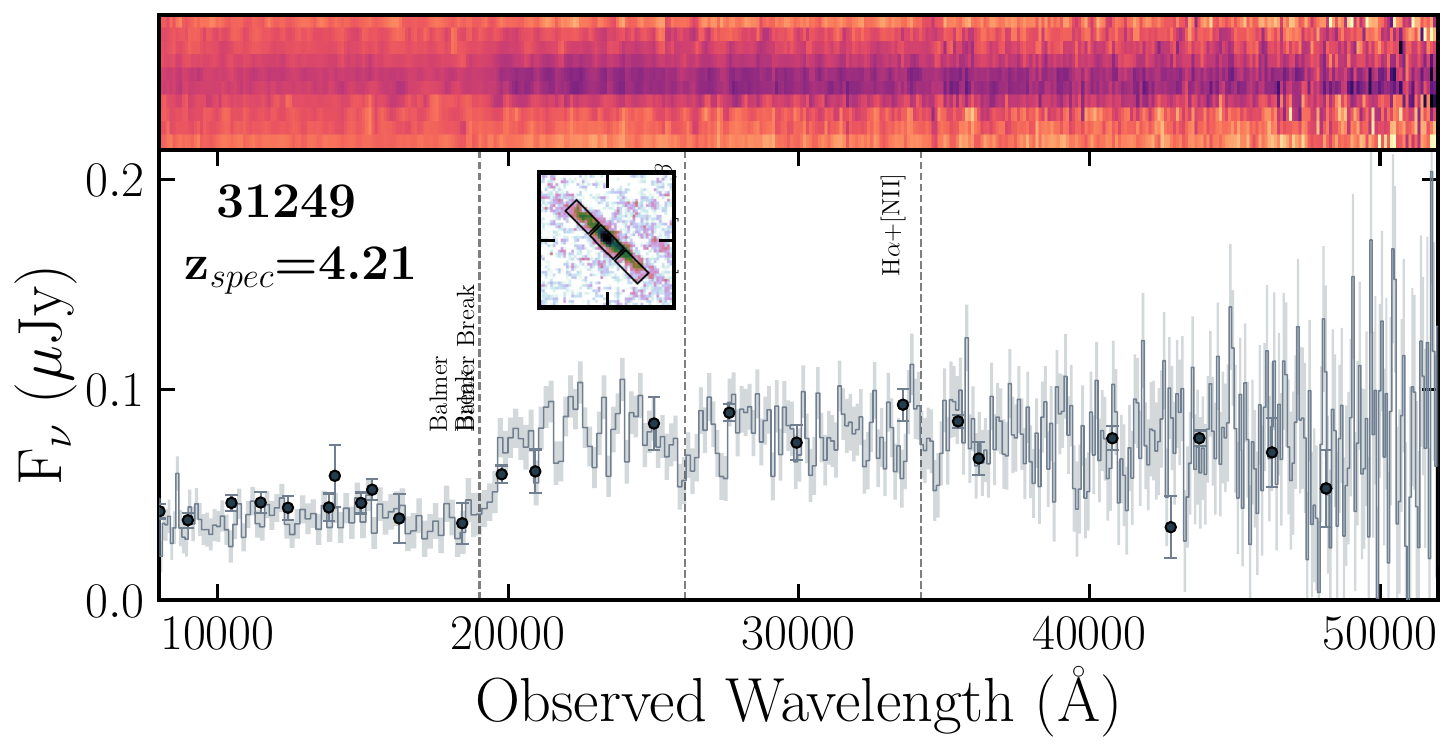}
\end{minipage}
\begin{minipage}[t]{0.48\textwidth}
    \centering
    \includegraphics[width=\linewidth]{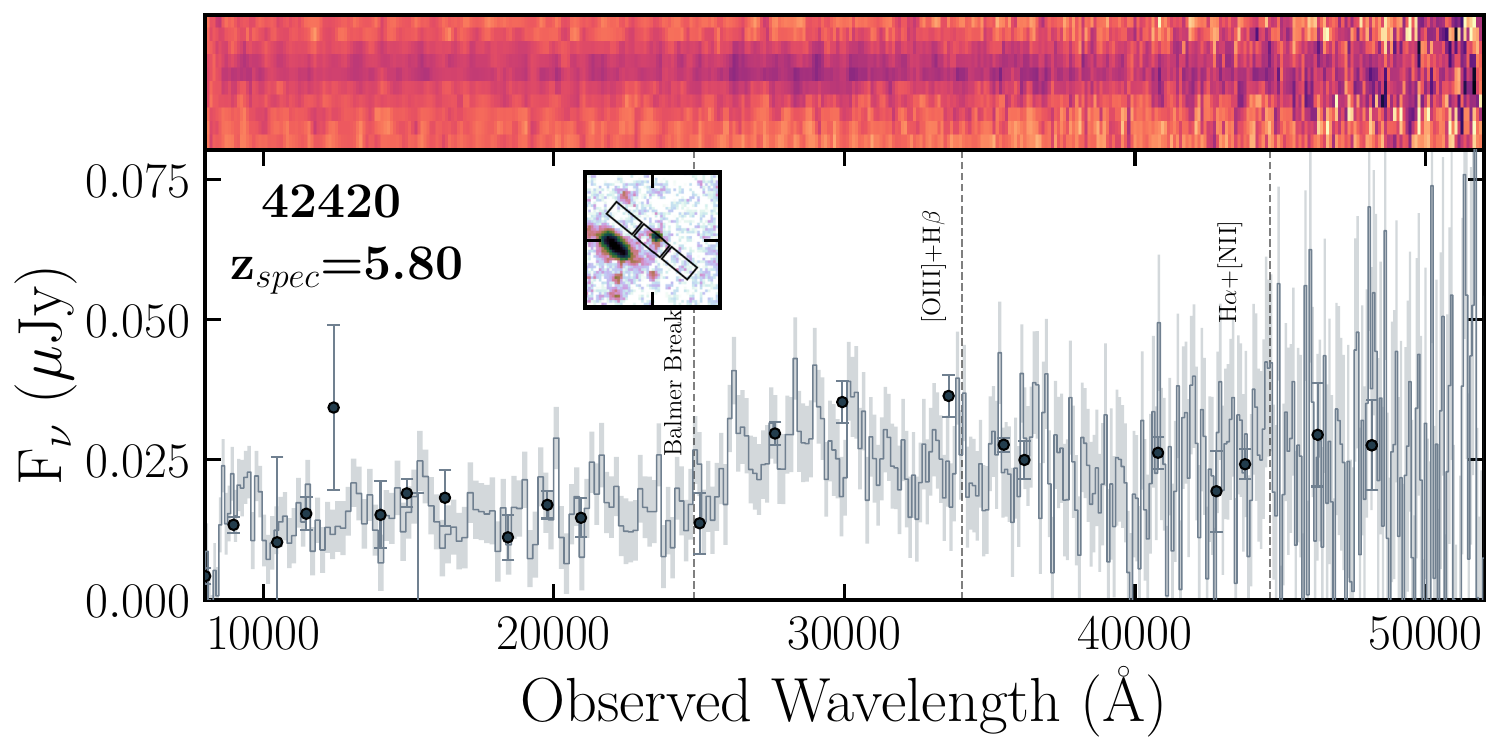}
    \vskip-5mm 
    \includegraphics[width=\linewidth]{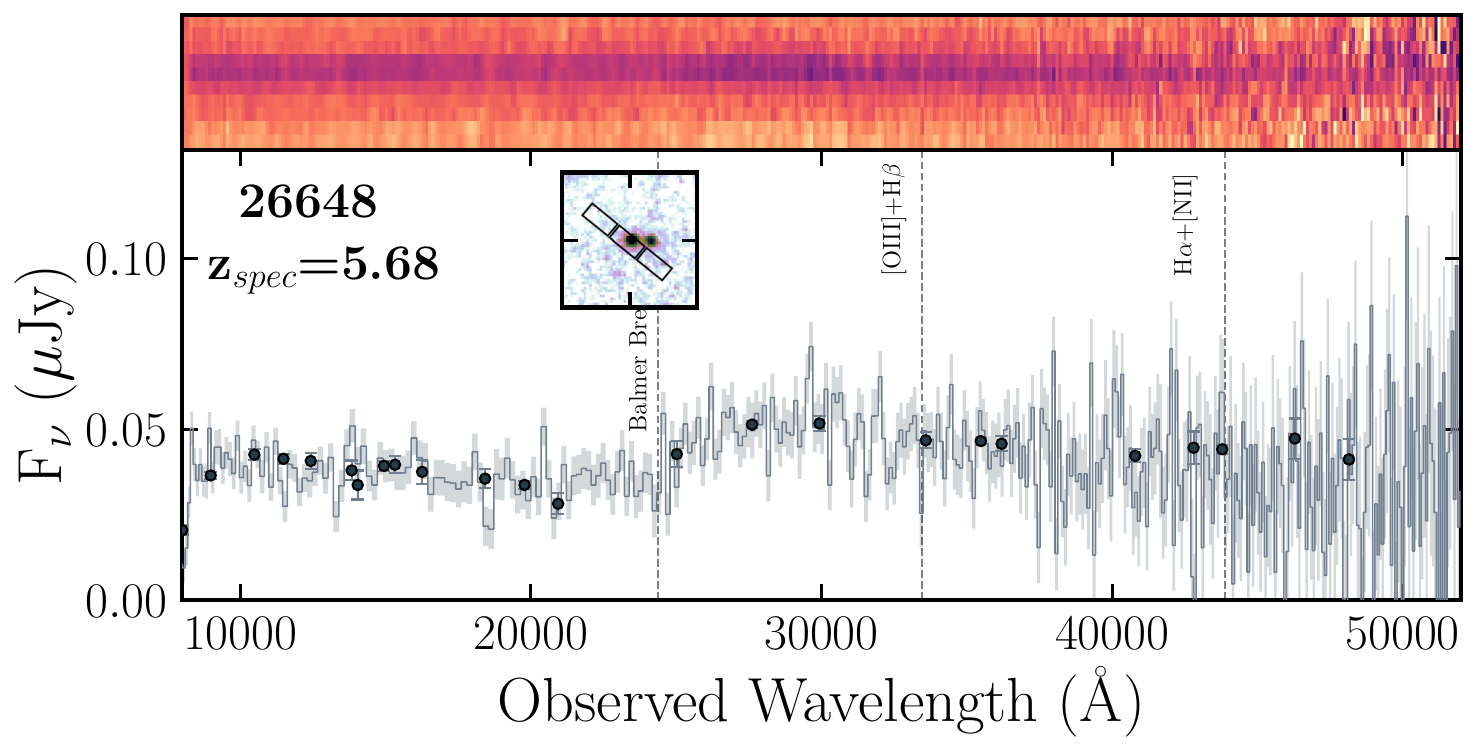}
    \vskip-5mm
    \includegraphics[width=\linewidth]{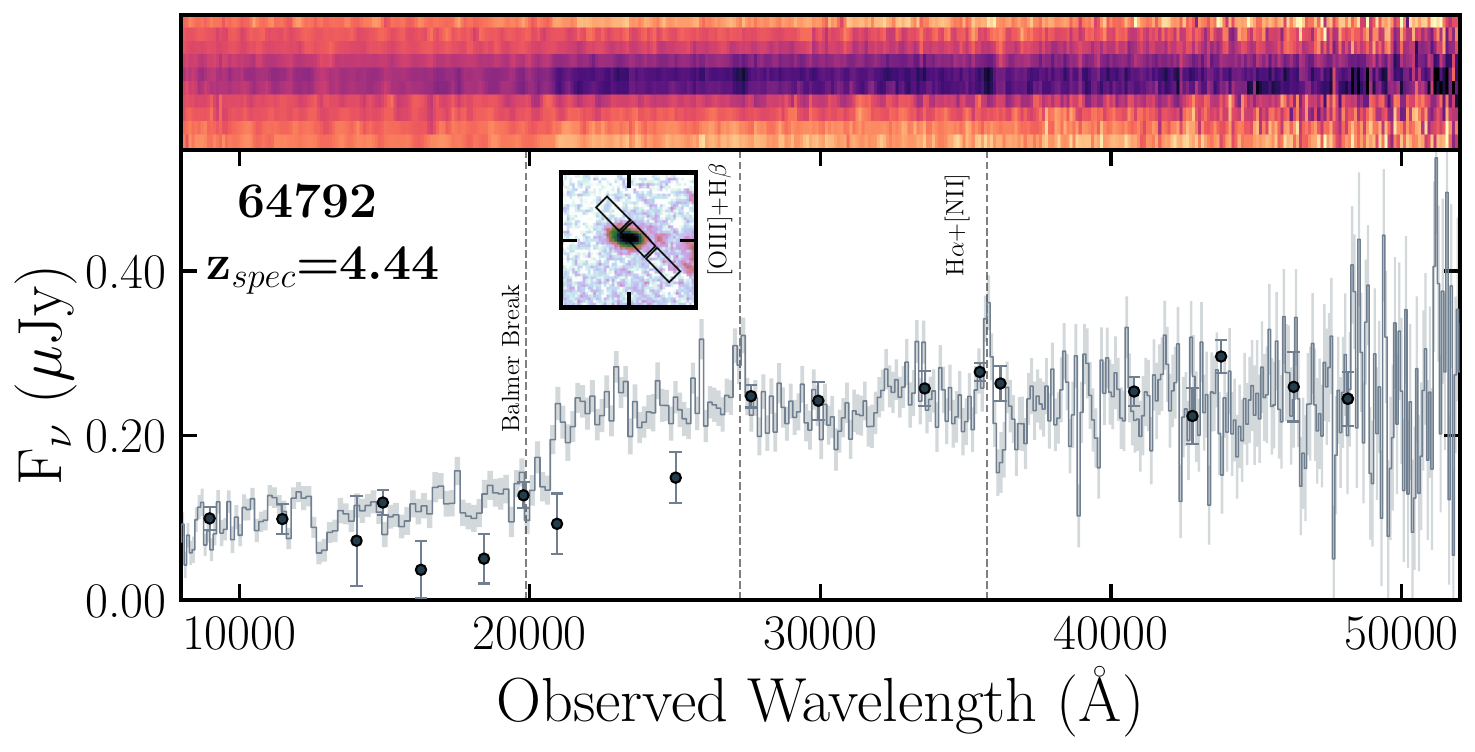}
    \vskip-5mm
    \includegraphics[width=\linewidth]{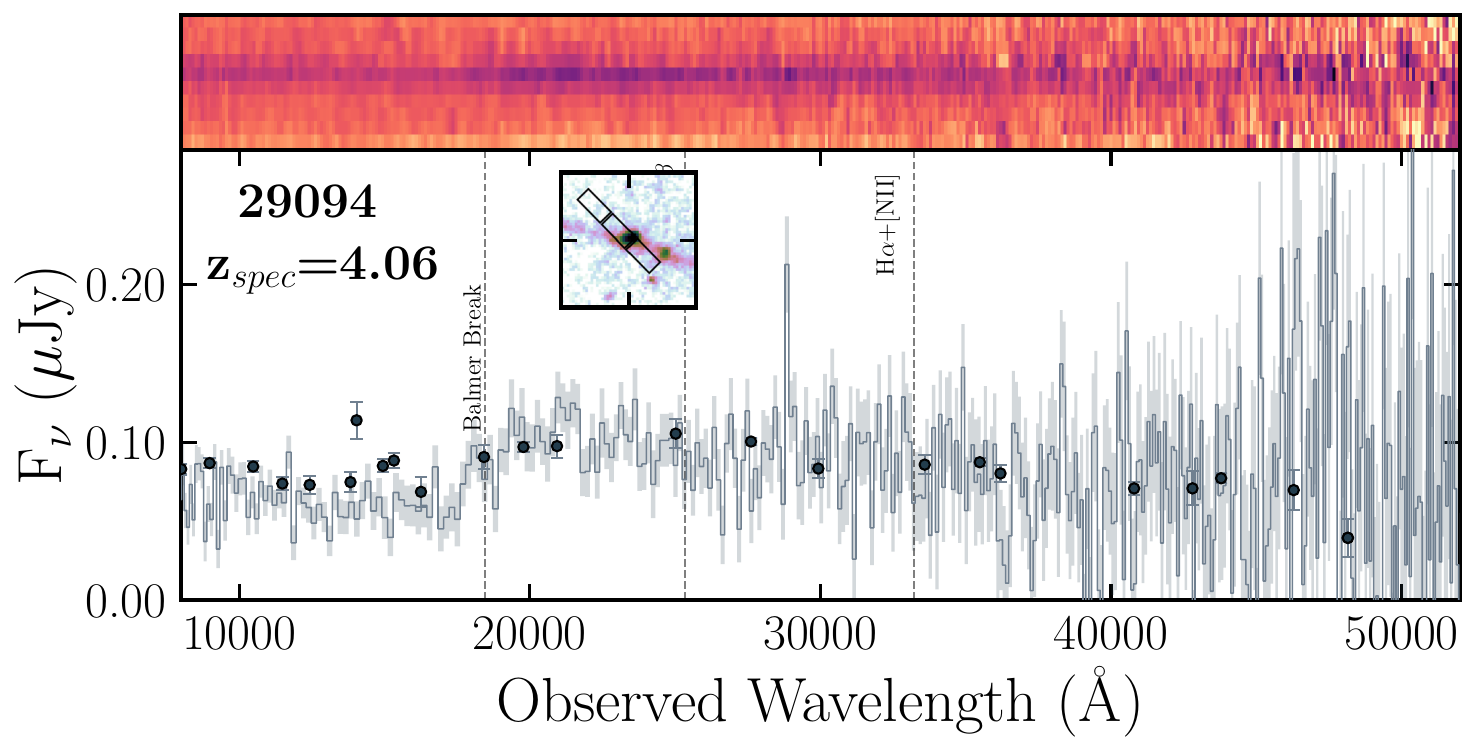}
\end{minipage}
\caption{\textbf{UNCOVER \textit{JWST}/NIRSpec PRISM spectra of the mini-quenched/napping galaxies at $\mathbf{4<z<7}$.} For each galaxy (in decreasing redshift) we show 2D (top panels) and 1D extracted spectra (bottom panels, black, in $F_{\nu}$ units) along with \JWST/NIRCam broad- and medium-band photometry (blue symbols, 0\farcs 32 apertures; \citealt{weaver2023, Suess2024}). Key spectral features are highlighted by dotted vertical lines. The \JWST/NIRCam F277W image for each galaxy is shown as an inset, overlayed with NIRSpec/MSA apertures. Each spectrum shows a Balmer break, weak or undetected nebular emission lines, and a slight UV upturn. Together, these spectral signatures point towards a recent downturn in star formation to low ongoing SFRs.}
\label{fig:spectra_gallery}
\end{figure*}

\newcommand{\sedcaption}{Free Parameters in the Fiducial \code{Prospector} Analysis}
\begin{deluxetable*}{ccc}
\tablecolumns{3}
\tablewidth{0pt}
\tablecaption{\sedcaption}
\tablehead{Parameter & Description & Priors}
\startdata
$M_{\rm total} (M_{\odot})$ & Total stellar mass formed & Log$_{10}$ Uniform: [10$^{7}$, 10$^{13}$] \\
$z$ & Redshift & TopHat: [$z_{phot}-0.25$, $z_{phot}+0.25$]\\
$Z/Z_{\odot}$ & Stellar metallicity in units of $Z_{\odot}$ & TopHat: [0.01,2]\\
$\tau_{\lambda, 2}$ & Diffuse dust optical depth & TopHat: [0.01, 3.00]\\
dust index & Power law index of the attenuation curve & TopHat: [-1.0, 0.4]\\
$\mathrm{spec}_{\rm norm}^{*}$ & Chebyshev polynomial; to scale the spectrum to match photometry & Order 5\\
$\mathrm{spec}_{\rm jitter}^{*}$ & multiplicative noise inflation term & TopHat: [0.5, 15.0]\\
$\sigma_{v}$ & Velocity dispersion (km/s) & Log$_{10}$ Uniform: [50, 500]\\
$\sigma_{v, emission}$ & Total velocity dispersion (Instrument + Emission Line, km/s) & Log$_{10}$ Uniform: [100, 500]\\
$t_{last}$ & Width of youngest age bin (Gyr) & TopHat: [0.01, 0.2$\times$Age of Universe at z$_{spec}$]\\
$\log(\mathrm{SFR}_{\rm ratio})$ & Ratio of the SFRs in adjacent age bins & Student-T: mean=0.0, scale=0.3, $\nu=1$\\
\enddata
\end{deluxetable*}
\label{table:sed_model}

The outline of this paper is as follows. In Section~\ref{sec:data}, we describe the spectroscopic and photometric dataset used in this work, while Section~\ref{sec:search} outlines the methodology to select and study napping galaxy candidates in the spectroscopic UNCOVER survey, and the medium-band MegaScience survey. Section~\ref{sec:census} describes the properties of nappers and their population statistics, and Section~\ref{sec:discussion} discuss them in the context of populations of mini-quenched/napping galaxies. Finally, we summarize our findings in Section~\ref{sec:summary}. Magnitudes used in this work have been calibrated with respect to the AB photometric system \citep{oke1983}. The fiducial cosmology model used for all distance measurements as well as other cosmological values assumes a standard flat cold dark matter universe with a cosmological constant $(\Lambda$CDM), corresponding to WMAP9 observations \citep{Hinshaw_2013}.

\section{Data and Methodology} \label{sec:data}

\subsection{UNCOVER/MegaScience photometry and spectroscopy}

We use \emph{JWST}/NIRCam \citep{rieke2022} UNCOVER \citep{Bezanson2024} and MegaScience \citep{Suess2024} Survey data, and ancillary datasets in Abell 2744 from \emph{HST} \citep{lotz:16,steinhardt:20,mahler:18} and other \emph{JWST} programs -- GLASS \citep{treu:22}, DDT observations (\#2756, PI: Chen). The UNCOVER NIRCam mosaic includes 7 filters (F115W, F150W, F200W, F277W, F356W, F444W and medium band F410M) with 5$\sigma$ depths of $\sim29.1$ AB mag at F444W. In this work, we use the 0.32'' circular aperture-based photometric catalogs constructed from PSF-homogenized mosaics (henceforth referred to as D032, v5.2.0) since all mini-quenched galaxies in this sample are compact sources; see \citep{weaver2023} for a detailed description of the aperture photometry pipeline). The medium-band MegaScience survey (PI: Suess; \citealt{Suess2024}) imaged the same footprint as UNCOVER in 13 additional bands (NIRCam F070W and F090W broadbands and F140M, F162M, F182M, F210M, F250M, F300M, F335M, F360M, F430M, F460M, and F480M medium bands). The reduction of the imaging for UNCOVER and MegaScience is detailed in \citet{Bezanson2024} and \citet{Suess2024}, respectively \footnote{All imaging observations used to create photometric catalogs used in this work can be found in MAST: \dataset[https://doi.org/10.17909/1ms3-sr76]{https://doi.org/10.17909/1ms3-sr76}. All photometric catalogs reside at the following DOIs: \dataset[https://zenodo.org/records/11059273]{https://zenodo.org/records/11059273}.}

All spectroscopic targets were observed with the \emph{JWST}/NIRSpec Micro-Shutter Assembly (MSA) follow-up of the UNCOVER/\JWST\ field Abell 2744 \citep{bezanson2022}. The spectroscopic experimental design and reductions are laid out in detail in \cite{price2025} \footnote{All spectra can be found in MAST: \dataset[https://doi.org/10.17909/8k5c-xr27]{https://doi.org/10.17909/8k5c-xr27}.} In this work, we use all UNCOVER PRISM spectra with confirmed spectroscopic redshifts and a median SNR per pixel of $>$ 2, resulting in 201 spectra at $z_{spec} > 1$, with 52 spectra at $4 < z_{spec} < 7$. For a comparative analysis of $z>3$ nappers published in the literature \citep{Looser2023a, Looser2023b, Strait2023, covelo2025}, we use publicly available PRISM spectra via the DAWN JWST Archive (DJA \footnote{https://zenodo.org/records/8319596} \citealt{heintz2024, degraaff2025}).

We use version (\texttt{v1.1}) of the \citet{furtak:22} parametric strong lensing model of Abell~2744\footnote{The \texttt{v1.1} lensing maps are publicly available at \url{https://jwst-uncover.github.io/DR1.html\#LensingMaps}}, similar to \citetext{Greene et al. 2023, Goulding et al. 2023, Wang et al. 2023, Atek et al. 2023}. We compute magnifications and their uncertainties for our sample at each object's position and spectroscopic redshift, primarily to correct for lensing-dependent parameters, namely stellar masses and star formation rates -- we use magnifications as a multiplicative correction factor.

\subsection{Stellar Population Synthesis Modeling with \code{Prospector}}

We conduct stellar population synthesis (SPS) modeling via Bayesian SED fitting and infer individual star formation histories and identify recent dips in star formation. Because we aim to explore the fidelity with which we can identify such sources given a variety of data characteristics, we perform this analysis on three datasets. First, we model the full NIRCam plus NIRSpec spectrophotometric data for spectroscopic targets observed at $4<z_{phot}<7$. Next we analyze the full UNCOVER/MegaScience photometric sample in the same redshift range (first fitting the full 20-band NIRCam photometric SEDs, and finally the broad-band only SEDs). The details of our modeling methodology are as follows.

For this analysis we utilize flexible star formation history modeling with the SED fitting framework \code{Prospector} \citep{Johnson2017, Leja2017,Johnson2021}. \code{Prospector} uses the Flexible Stellar Population Synthesis (FSPS) stellar population synthesis models \citep{Conroy2009, Conroy2010}, the MILES spectral library \citep{Sanchez-Blazquez2006,Falcon-Barroso2011}, and the MIST isochrones \citep{Choi2016, Dotter2016}. We use non-parametric SFHs and implement a flexible age bin model, which utilizes fixed time/age bins at early times (with continuity priors on the ratio of SFR in adjacent age bins), flexible bins that each form the same amount of total stellar mass at intermediate times, and a final age bin with a flexible age boundary (closest to the epoch of observation); this model is optimized to constrain mass assembly in recently quenched galaxies (see \citealt{Suess2022b,setton2023}). This scheme is designed to recover quenching timescales and burst mass fractions with spectro-photometric data \citep{Suess2022b, Suess2022a}. 

In our fiducial model, we define the 2 fixed time bins at early times, such that they sample 40\% of the age of the Universe (0.4 $\times$ t$_{univ}$), while simultaneously fitting for the redshift, z$_{spec}$. The youngest age bin is sampled in the analysis in the range [0,0.3 $\times$ (t$_{univ}$)], as well as the flexible age bins. 

We assume a \cite{Chabrier2003} initial mass function. We adopt the \cite{Kriek2013} dust law with $A_v$ and dust index as free parameters, with doubled attenuation around young ($<10^7$ yr old) stars following \cite{Calzetti1997,Wild2020,Suess2022a,setton2023}. 
Our fiducial model is also agnostic to the physical mechanisms causing line emissions, i.e., we implement the nebular line marginalization framework within \code{Prospector}.
Finally, we also include as free parameters the stellar and gas-phase metallicity (independent of each other). This analysis does not include any priors derived from any mass-dependent priors \cite{Behroozi2019,setton2023}, as the SED fitting is conducted in the image plane of A2744, without any strong lensing magnification-correction (which is implemented post-SED fitting). This is distinct from the estimation of stellar population properties and photometric redshifts with \code{Prospector}-$\beta$ \citep{prospbeta2023}, where the magnification is being varied simultaneously with the stellar population parameters. We use the \texttt{dynesty} dynamic nested sampling package \citep{Speagle2020} to sample the posterior distributions.   

There are minor differences in our fiducial model when fitting spectra vs photometry-only observations. For spectrophotometric fitting, we use a fifth order Chebyshev polynomial as a relative and absolute flux calibration vector, to marginalize over any systematic wavelength-dependent differences in flux between the photometry (which we assume has the most robust flux calibration) and the observed spectra. We fit for a spectroscopic redshift by sampling the range [$z_{spec}$ - 0.05, $z_{spec}$ + 0.05], to allow for an extended posterior distribution for the redshift inferred from spectroscopic redshift measurement from an emission line and/or flux break. For photometric fits, we use the range [0.5,12] as a prior for the redshift parameter. Moreover, in photometry-only fitting, since we keep the redshift parameter free, the flexible star formation history bins adapt the time available to form stars with the age of the Universe corresponding to the redshift value in the fitting process. 

Free parameters in this model are listed in Table \ref{table:sed_model}. 

\begin{figure*}[htb!]
\centering
\includegraphics[page=5,width=1.0\textwidth]{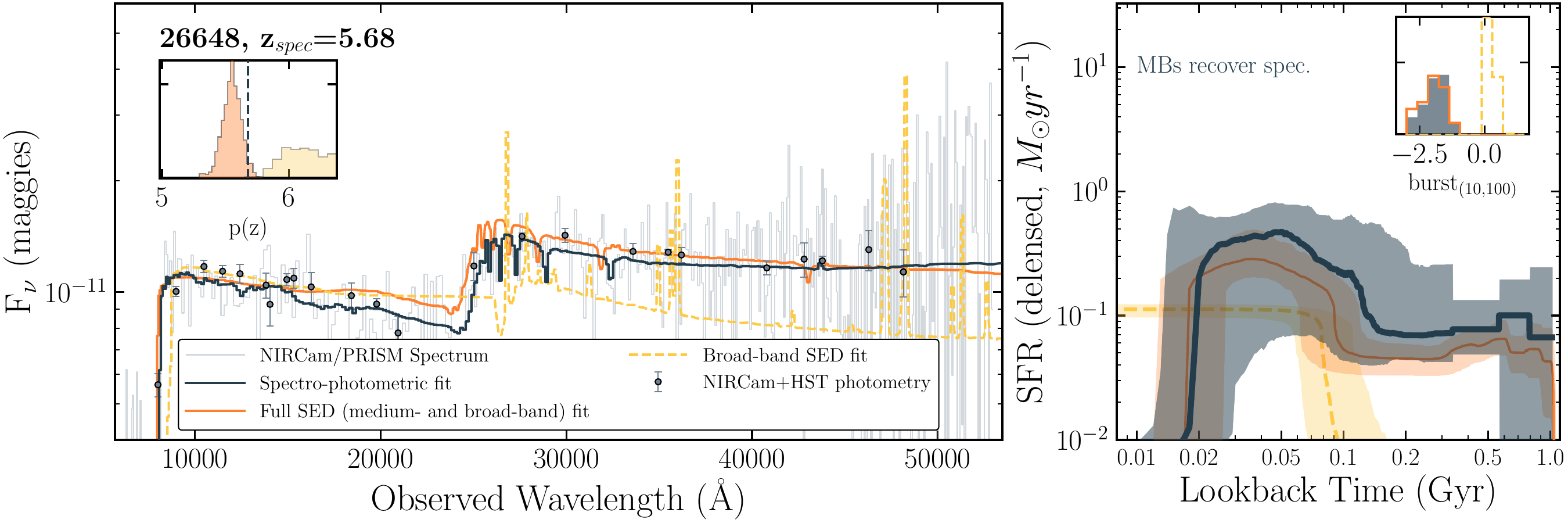}
\includegraphics[page=3,width=1.0\textwidth]{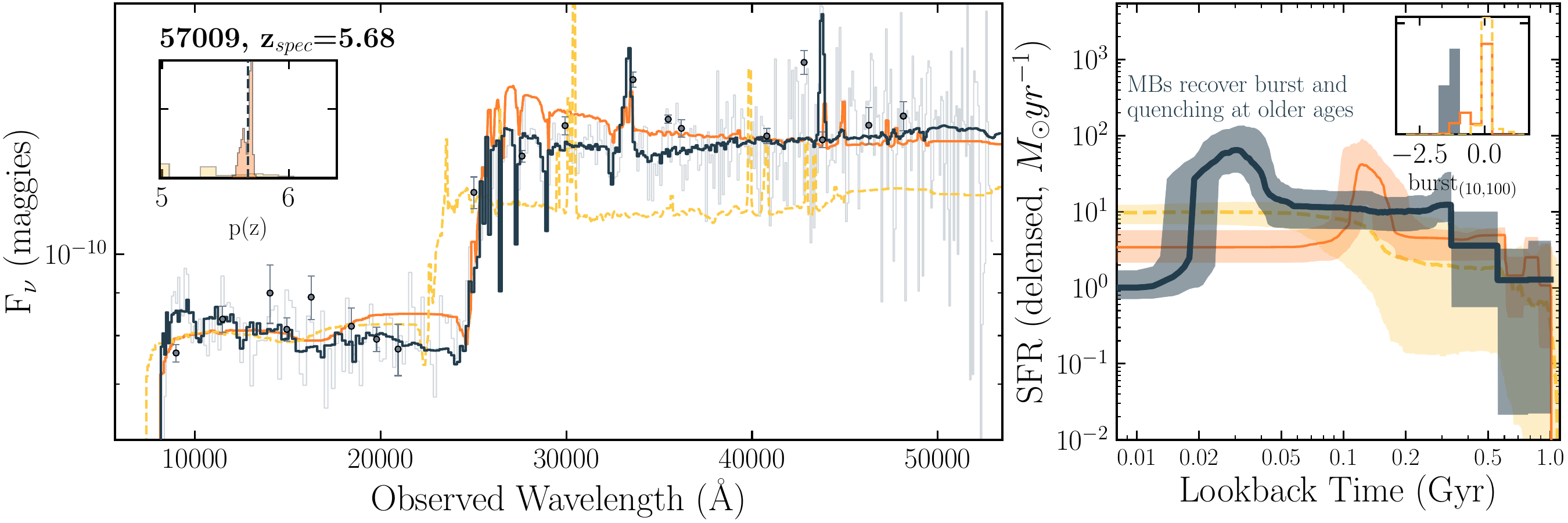}
\includegraphics[page=9,width=1.0\textwidth]{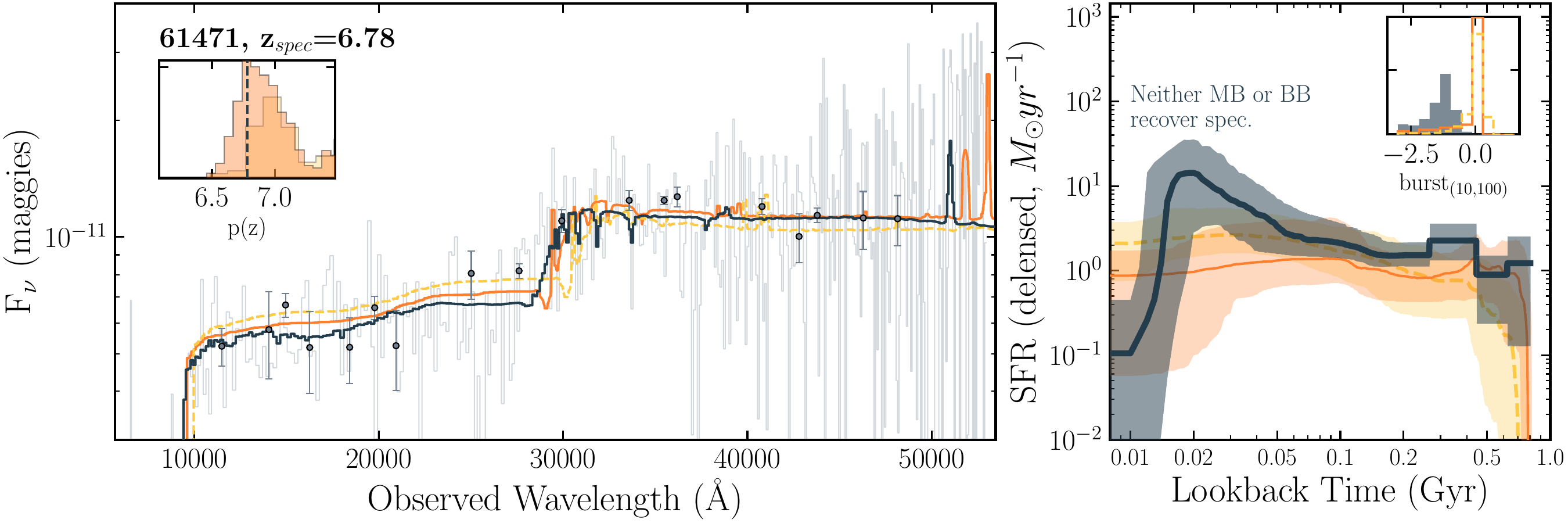}
\caption{\code{Prospector} spectrophotometric modeling of three representative examples for which spectrophotometric and photometric-only models strongly/marginally agree (top/middle row) and disagree (bottom row). \textbf{(Left)} the NIRSpec/PRISM spectra and NIRCam photometry are shown with maximum a posteriori (MAP) models from fits to three subsets of the data: full spectrophotometric fits (dark blue), full photometric SED with broad- and medium- bands (orange), and broadband-only fits (yellow); the retrieved \textbf{(Right)} SFH posterior distributions (delensed), with the burstiness parameter ``burst$_{(10,100)}$" (log(SFR$_{10}$/SFR$_{100}$)) for each model in the inset. In all cases, SPS modeling of the broadband photometry is insufficient to recover late time dips in star formation (or non-zero burstiness) needed to identify and characterize nappers, in part due to redshift uncertainties (e.g., UNCOVER 61471). Some MB fits (e.g., UNCOVER 26648) recover the spectroscopic fit-based SFH, including the burst and quenching episode. In cases like UNCOVER 57009, MB fits recover burst and quenching, but at epochs different from that of the spectroscopic fits. In UNCOVER 61471, the spectrophotometric fits infer a rising SFH (500 to 200 Myr before observation) and a quenching episode ($\sim$200 Myr before observation); the MB fits do not robustly recover these, and differ in SFRs by $\sim 1\sigma$.} 
\label{fig:burstiness_2}
\end{figure*}

\section{How do we find Nappers?}\label{sec:search}

Our ultimate goal is to use the small spectroscopic samples of napping galaxies to extend to photometric identification. This Section details our selection methodology. We show that SFHs derived from photometric fits -- even those with extensive medium bands -- do not consistently capture recent SFR downturns(described in \S3.1). Therefore we define empirical spectral indicators that uniquely identify napping galaxies (\S3.2), apply those selections to the UNCOVER/MegaScience broad- and medium-band photometric catalog (\S3.3), and validate this targeting methodology using an expanded spectroscopic sample (\S3.4). With this approach, we construct a robust sample of photometric nappers and constrain their number densities. In this study, we are calibrating our intuition of spectroscopic signatures and SFHs of nappers, and building spectroscopy-based indicators for datasets where only broad- and mediumb-band photometry exist. 

\subsection{Calibrating empirical identification with spectrophotometrically-derived star formation histories}

A handful of napping galaxies were targeted by the UNCOVER NIRCam/PRISM spectroscopic sample. We collect a sample of these spectra for calibration purposes through a combination of visual inspection and SPS modeling, emphasizing that we aim for purity not completeness.

From 52 UNCOVER spectra at $4<z<7$, we visually identify systems that exhibit a visible Balmer break (and remove any spectra from the parent sample that have spectral reduction artifacts). We fit \code{Prospector} models to these spectra -- along with broad-band (BB) and medium-band (MB) photometry -- and isolate galaxies where SFHs contain a recent SF burst ($\sim$100-200 Myr in lookback time) and a recent quenching episode ($\sim$50-100 Myr in lookback time).

The sample of such systems -- UNCOVER nappers -- is shown in Figure \ref{fig:spectra_gallery}, in order of decreasing redshift. Each 2D and 1D spectrum is shown, along with the full photometric SED. F277W images of each is shown as an inset, with the MSA aperture annotated. The sample is identified by the characteristic SED shape, including significant Balmer breaks - indicating at least temporary diminished star formation rates, residual UV flux and minimal H-$\alpha$ and [OIII] emission, which are the hallmarks of napper spectra published in literature \citep{Looser2023a, Looser2023b, Strait2023, covelo2025}. The sources exhibit a range of morphologies -- from compact and round to extended disks.

We show representative model outputs for three spectroscopic targets in Figure \ref{fig:burstiness_2}. Each row shows the spectrophotometric data (in gray) with best-fitting models (in color) in the left panel and recovered star formation histories in the right panel. On the left, the redshift probability distribution is shown as an inset, with the spectroscopic redshift indicated by a vertical dashed line. Finally, we include posterior distributions of the ratio of SFRs averaged over two timescales to identify late-time dips, comparing short (averaged over 10 Myr) to longer (100 Myr) timescales with $SFR_{10/100}$ -- this is the nominal estimate for ``burstiness'' in the literature (\citealt{Looser2023b}, \citealt{cole2023}). In each panel, fits to the full spectrophotometric data are indicated by thick bands (to include uncertainties), fits to the full medium and broad-band photometry in orange, and to only the broad-band photometry in yellow. 

\begin{figure}[htbp]
\includegraphics[width=0.5\textwidth]{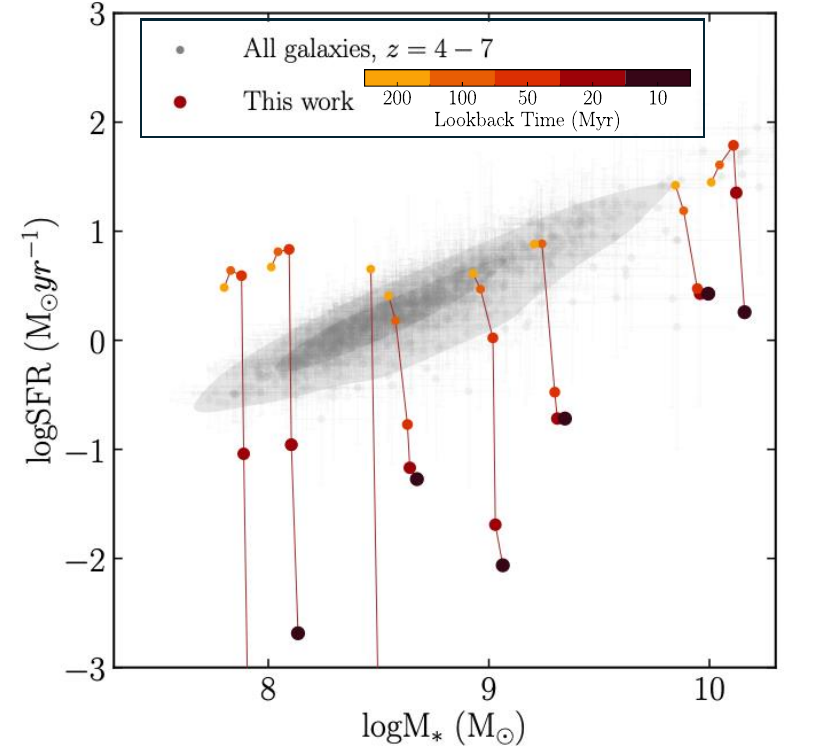}
\caption{The star formation rates averaged over a range of timescales (10-200 Myr) versus stellar mass for the spectroscopic napper sample. Grey points show the SFR$_{10\,Myr}$ for the parent photometric sample at $4<z_{phot}<7$. Each individual track shows a galaxy moving from on/above the main sequence (MS) to significantly below the MS, as expected from post-starburst galaxies. The "horizontal" movement of SFR values for a given galaxy is only approximate, for the purposes of clarity. SFRs plotted outside the y-limit of the plot are consistent with zero.}\label{fig:burstiness}
\end{figure}

In both examples, the spectrophotometric fits clearly recover decreased late-time star formation. We quantify this statement for the full sample by integrating the SFR ratio posteriors -- for the majority of the galaxies, the probability distribution of $SFR_{10/100}$ falls at least $>2\sigma$ away from 0 (corresponding to constant SFR). We also show this evolution in SFR versus stellar mass for the spectroscopic sample in Figure \ref{fig:burstiness}. Ideally, one would hope that the full collection of medium and broad-band photometric data would provide sufficient sampling of the SED to robustly identify these higher order SFH fluctuations (orange). For 4 out of 8 spectroscopically verified nappers, the full photometric SED robustly identifies the SFR dip, as exemplified by UNCOVER-57009 in the middle row of Figure \ref{fig:burstiness_2}.  However, there are 4 spectroscopic nappers for which the photometric data (BB + MB) fail to capture late-time evolution, either by not recovering a burst/quenching episode, or estimating burst/quenching at a time that is in disagreement with the spectroscopic inference. The bottom row highlights UNCOVER-61471, for which the photometric modeling infers steady star formation over the last 100 Myrs. UNCOVER-61471 has the highest redshift; at $z=6.78$, H$\alpha$, which is generally used as an instantaneous SFR indicator, is observed at 5.1$\mu$m -- it falls beyond the reddest NIRCam filters, but is at the edge of the NIRSpec/PRISM spectrum (only spectrscopic napper not sampling H$\alpha$ with NIRCam photometry). Although relatively noisy at the red end, the spectrum provides critical information about H$\alpha$ and weaker emission features, which subsequently allows us to constrain the continuum precisely. We note that the broadband SEDs are unable to identify late time SFR deviations (yellow).

We show spectrophotometric SED fits and SFHs for the remaining 5 galaxies in Appendix \ref{sec:sed_fits_appendix}. 
To quantify the burstiness of UNCOVER nappers, we measure the distributions of SFRs as a function of timescale based on SFH constraints, shown in the logSFR-logM$_*$ parameter space (see Figure \ref{fig:burstiness}). We also show SFR measurements (averaged over 10 Myr) for UNCOVER galaxies -- at $4 < z < 7$ (constrained via photometry-only fitting), where contours demonstrate the logSFR-logM$_*$ correlation at these redshifts. SFR measurements averaged over varying timescales -- 10, 20, 50, 100 and 200 Myr -- for the spectroscopic systems clearly show that for each galaxy, the SFR as a function of time first increases rapidly and then sharply decreases closer to the epoch of observation, as expected from galaxies undergoing stochastic SFH and recent abrupt quenching (or observed in the quenched phase). Although there is no universal definition of quiescence in the literature, there is a consensus that such a definition should be relative and evolve with redshift as the average sSFR(z). Such definitions can either scale with timescales (e.g., a fraction of the Hubble time) or be made with respect to the evolving ''star formation main sequence'' (e.g., 2-3 $\sigma$ below the average relation) \citep[][and references therein]{whitaker2026}. 

\subsection{Quantifying identification with spectroscopic indicators}\label{sec:findnappers}

Based on our inference of star formation histories of UNCOVER galaxies at $z=4-7$, we identify critical spectral features corresponding to nappers. We design a quantitative criterion to identify these galaxies via medium- and broad-band photometry (with modeling that is agnostic to physical SED models), as they are not isolated in classic UVJ \citep{williams2009} and $(ugi)_s$ \citep{Antwi-Danso2022} color-color spaces (see a detailed discussion on this in Section \ref{sec:mb_cat}). 

Extant studies take three primary approaches to isolate post-starburst systems like nappers: (a) by constraining spectral features (Balmer break, equivalent width of H$\alpha$, etc.) in spectroscopy and photometry based on SPS modeling and SED fitting (e.g., \texttt{BAGPIPES} maximum a posteriori -- MAP -- models in \citealt{covelo2025}), (b) relying on substitutes/proxies for napper-unique features using observed magnitudes/fluxes and colors (e.g., $F335M-F356W$ vs $F410M-F444W$ color spaces in \citealt{trussler2024}), or (c) a mix of the two approaches (e.g., \citealt{Looser2023b}).  In this work, we attempt to stay as close to the empirical measurements as possible. As described and demonstrated in \cite{mitsuhashi2026}, physically-motivated SPS models at high redshifts may have inferred properties that are likely biased, due to various factors, including incorrect photometric redshifts, model emission lines (and ratios) not calibrated to high-redshift galaxy samples, and/or a heterogeneous sampling of spectral features by broad- and medium-band photometry.

\begin{figure*}[htb!]
\centering
\includegraphics[width=1.0\textwidth]
{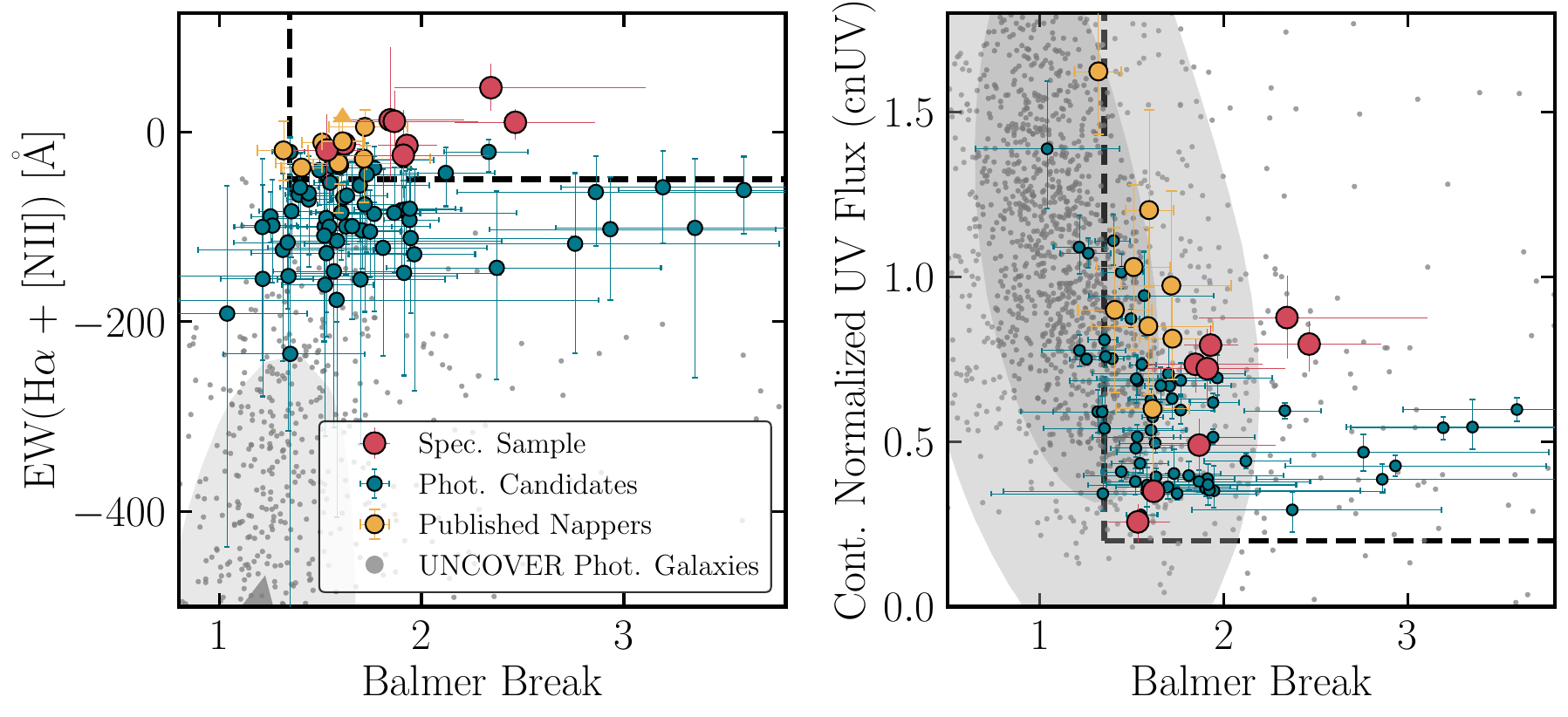}
\caption{  \textbf{Photometric identification of nappers using Balmer break, H$\alpha$ equivalent widths and (optical) continuum-normalized UV flux} in UNCOVER/MegaScience medium-band photometric and SED catalogs \citep{Suess2024,wangsps2024} (grey). These estimates have been performed via empirical Bayesian model fitting to medium+broad band photometric SEDs. \textbf{(Left)} Equivalent width (H$\alpha$+[NII]) vs Balmer Break measurements for the napper sample in this work. The cuts on EW(H$\alpha$+[NII]), in angstroms) and Balmer break strength are marked with black solid lines; all photometric candidates within 1$\sigma$ of the designed thresholds along any of the three parameters are included in our sample (60 objects make this cut, while 15 nappers have 50th percentile of EW(H$\alpha$+[NII] greater than the threshold). Spectroscopically-confirmed nappers in this study are labeled red, while medium+broad band-selected nappers are in teal.  \textbf{(Right)} UV color vs Balmer break strength for the same sub-populations. Published galaxies from \loosera, \looserb and \strait are plotted in orange; the \loosera system does not have an H$\alpha$ emission measurement. Nappers are significant outliers -- they lie at the extreme end of the full galaxy population at these redshifts, with strong Balmer breaks and weak/absent H$\alpha$, and bluer UV colors for the majority of the sample.}
\label{fig:nap_parameter_space}
\end{figure*}

We construct the following indicators. 

\begin{enumerate}

\item \textbf{Equivalent Width (EW) of H$\alpha$+N[II]}: In PRISM spectra, H$\alpha$+N[II] lines are blended. Hence, our first indicator is EW(H$\alpha$ + N[II]), used as a proxy for the instantaneous specific star formation rate (sSFR) in literature (or sSFR averaged over the most recent 10-20Myr in SFH estimation via SPS modeling; \citealt{Fumagalli2012,wanglilly2020,Khostovan2021}). The correlation between EW(H$\alpha$) (unblended) and sSFR arises from the consideration that the line strength of (H$\alpha$) estimates the instantaneous SFR, while the continuum estimate in the rest-frame optical correlates with the remnant stellar mass of the galaxy. We compute the equivalent width (EW) over $[\lambda_1,\lambda_2]$ relative to the local continuum $F_{\lambda,\mathrm{cont}}$ as

\begin{equation}
\mathrm{EW} \equiv \int_{\lambda_1}^{\lambda_2}\left(1-\frac{F_{\lambda}}{F_{\lambda,\mathrm{cont}}}\right)\,\mathrm{d}\lambda,
\label{eq:equivalent_width}
\end{equation}

where $F_{\lambda}$ is the observed flux density. With this sign convention, absorption features have $\mathrm{EW}>0$, while emission features have $\mathrm{EW}<0$. We use the range [6230,6700]\AA\ as [$\lambda{_1}$, $\lambda{_2}$]. To estimate $F_{\lambda,\mathrm{cont}}$, we fit a line using the spectra bluewards ([6230,6430]\AA) and redwards ([6750,6950]\AA) of the blended emission lines, while the lines are considered to be sampling the range [6430,6700]\AA. A low value of this estimator indicates no/weak star formation in the recent epoch (see Equation \ref{equation:cuts} below). 

\item \textbf{Balmer break strength}: The amplitude of the Balmer break, measured straddling $\sim$3650\AA\, is our second estimator. A strong break typically indicates a stellar population dominated by A-type stars with deep Balmer absorption lines.  This is a signature that emerges once short-lived O/B stars have died off following the cessation of star formation, typically peaking a few hundred Myr to $\sim$1 Gyr afterward (i.e. evidence of recent quenching; see \citep{mintz2025} and references therein).

We define the Balmer break as the ratio of the median observed flux density measured redward and blueward of the Balmer break:
\begin{equation}
    BB
    \equiv \frac{\widetilde{F}_{\nu,\,\mathrm{red}}}{\widetilde{F}_{\nu,\,\mathrm{blue}}}
    = \frac{\mathrm{median}\left[ F_\nu(4150\text{\AA} \le \lambda \le 4250\text{\AA}) \right]}
           {\mathrm{median}\left[ F_\nu(3400\text{\AA} \le \lambda \le 3600\text{\AA}) \right]} \, .
\end{equation}
where $\widetilde{F}_{\nu,\,\mathrm{red}}$ and $\widetilde{F}_{\nu,\,\mathrm{blue}}$ denote the median observed flux densities in the wavelength intervals [4150, 4250]\,\AA\ and [3400, 3600]\,\AA, respectively. A higher value, or a stronger Balmer break, is most strongly associated with recent quenching of star formation, peaking a few hundred Myr to $\sim1$ Gyr after cessation and declining thereafter, in contrast to Dn4000 , which increases roughly monotonically with age \citep{kauffman2003}.

\item \textbf{Optical continuum-normalized UV flux, or cnUV}: If the optical continuum used in the EW(H$\alpha$ + N[II]) estimation is a proxy for a napper's stellar mass and UV flux correlates with star formation at the 100-200 Myr timescale (in literature, this UV flux/luminosity is measured at 1800\AA), then continuum-normalized UV flux (cnUV) approximately correlates with the sSFR averaged over $\sim$100 Myr. 

Let \(F_{\nu,\,\mathrm{UV}}(1800\,\text{\AA})\) be the observed flux density at 1800\,\text{\AA}, and \(F_{\nu,\,\mathrm{cont}}\) be the optical continuum flux density (the same continuum -- but in flux density units -- used for EW(H\(\alpha\)+[NII])).

\begin{equation}
\mathrm{cnUV} 
\equiv 
\frac{F_{\nu,\,\mathrm{UV}}(1800\,\text{\AA})}
     {F_{\nu,\,\mathrm{cont}}}.
\end{equation}

A high value of cnUV indicates active star formation in the recent past. Note that we do not correct cnUV for dust attenuation; we explore the impact of dust correction in Appendix \ref{sec:cnuv_appendix2}.

\end{enumerate}

We isolate nappers with the following selection:

\begin{equation}
\begin{aligned}
&(\textbf{EW(H$\alpha$ + N[II])} > -65\mathrm{\AA})\ \wedge \\
&(\textbf{Balmer Break} > 1.35)\ \wedge \\
&(\textbf{cnUV} > 0.2)
\end{aligned}
\label{equation:cuts}
\end{equation}

Each indicator is lensing invariant, and relies on estimators analogous to sSFRs. Moreover, unlike EW(H$\alpha$) and Balmer break-based napper identification (e.g., in \citealt{covelo2025}), we use the additional indicator cnUV that removes (a) maximally quenched and/or extremely dusty (HST or NIRCam/SW- dark) galaxies to differentiate them from the post-starbursty SFHs of nappers, and (b) provides a constraint on the turnover in SF between 10-100 Myr prior to the epoch of observation.

All 8 spectroscopically confirmed UNCOVER nappers presented in this work fulfill the threshold presented in the previous section. As mentioned in Section \ref{sec:data}, we use UNCOVER spectra from DJA (v4.4), which are flux normalized to NIRCam photometry. Each spectrum (in F$_\lambda$ units) is converted to rest-frame wavelengths, before estimating the above indicators. In Figure \ref{fig:nap_parameter_space}, we show each UNCOVER napper in the \ewha vs Balmer break, and cnUV vs Balmer break parameter spaces, in comparison with all UNCOVER galaxies at $z=4-7$.

This parameter space also identifies published post-starburst/mini-quenched systems in literature -- the lower redshift ($z_{spec} = 4.5$) system from \citep{Looser2023b}, the post-starburst galaxy ($z_{spec} = 5.2$) reported in \citep{Strait2023}, as well as the $z_{spec}$=7.3 \citep{Looser2023a} napper (please note the JWST NIRSpec/PRISM spectrum does not sample rest-frame H$\alpha$ for this galaxy.) We also recover 5/10 nappers identified by \cite{covelo2025}; the differences in our selection methodology based on empirical measurements, vs SPS model-based measurements in \cite{covelo2025} are illuminated further in Section \ref{sec:comparisons}.

\begin{figure*}[htb!]
\centering
\includegraphics[width=1.0\textwidth]
{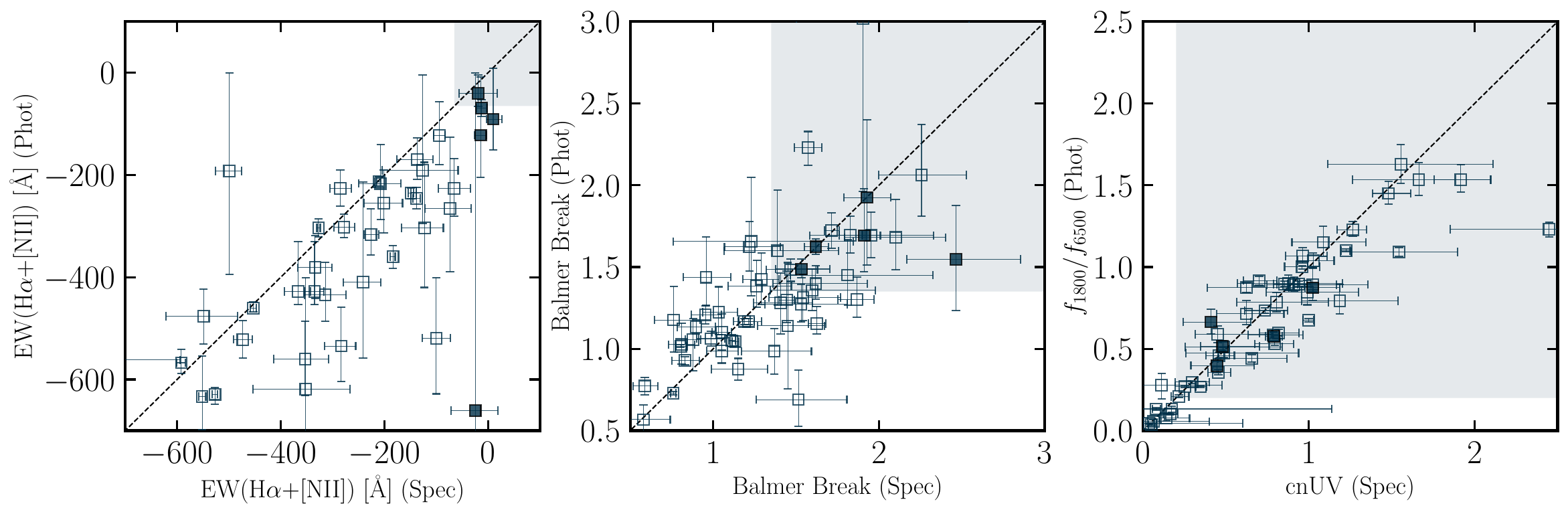}
\caption{ (Left) EW(H$\alpha$+[NII]),  \textbf{(Center)} Balmer break, and \textbf{(Right)} cnUV measurements from empirical Bayesian fitting to MB+broad band photometry, compared with empirical measurements from spectroscopy. The filled region is the parameter space corresponding to nappers. The photometric measurements recover spectroscopically confirmed nappers (filled squares) in all instances, for objects in the redshift range 4$<z<$6.6. No false positives are observed in this analysis, i.e. the sample of napper candidates is likely pure, albeit incomplete.}
\label{fig:spec_vs_emp}
\end{figure*}

\subsection{With medium-band (MB) photometry catalogs} \label{sec:mb_cat}

Following spectroscopic identification and calibration, we select nappers based on the rich photometric dataset in UNCOVER/MegaScience. Here, we use results from \cite{mitsuhashi2026}, who model SEDs without any dependence on a physical stellar population synthesis model (e.g., FSPS). To summarize, \cite{mitsuhashi2026} use a power law to model rest-frame UV and optical continua (and derive Balmer breaks from inferred flux ratios), while emission lines are formalized via a Gaussian model. These SED/spectra models are converted to fluxes using \texttt{sedpy} and fitted to observed photometry, with redshift priors borrowed from the photometric redshift posteriors from \cite{wangsps2024} (that employs the \texttt{prospector}-$\beta$ framework). We note that these model inferences depend on filter placement (and what they sample at a particular photometric redshift), filter size, as well as the intrinsic galaxy SEDs/properties (e.g., emission lines, continuum/line confusion, minimal underlying continuum emission, etc).

The Balmer break strength is estimated from the inferred models by taking the ratio between the average continua in the rest-frame UV and optical continua at 3700--3900\AA\ and 4150--4250\AA, respectively. The EWs are measured by dividing line luminosities by the continuum at the line wavelength, i.e., 6563\AA\ for H$\alpha$+[N\,{\sc ii}]. Moreover, we calculate flux at 1800\AA\ and 6500\AA\ as a substitute for rest-frame UV and optical continuum flux, to compare directly with the spectroscopic measurements of cnUV. 

Figure \ref{fig:nap_parameter_space} shows the photometrically selected nappers (in teal), with uncertainties calculated via convergence criteria and Monte Carlo sampling of posterior chains described in \cite{mitsuhashi2026}. We find 60 UNCOVER galaxies within 1$\sigma$ of the criterion discussed in Equation \ref{equation:cuts} at $z=4-6.6$ (15 nappers have 50th percentile of EW(H$\alpha$+[NII] greater than the threshold) -- these objects show best-fit models that are consistent with napper-like SEDs, with high \ewha\ (lower in absolute units, compared with line emitting galaxies), as well as relative strong Balmer break values, and high values of cnUV.


\subsection{Validating Napper Identification with Medium Band Photometry} \label{sec:recovery}

In Figure \ref{fig:spec_vs_emp}, we compare values of napper indicators EW(H$\alpha$+[NII]), Balmer break strength, and cnUV of all UNCOVER spectra in the redshift range 4$<z<$7 (52 robust spectra), from model-based fitting to MB+BB photometry in \cite{mitsuhashi2026}, with empirical measurements from spectroscopy. The DJA v4.4 archive contains 80 UNCOVER spectra (74 unique systems) between $z=4-6.8$ at average SNR $>$ 3 (which matches the PRISM wavelength coverage that allows us to reliably measure continuum redward of Halpha). 52 out of 74 spectra have reliable Bayesian MB+BB fits (selected for data in all available \JWST filters, F444 AB mag $<$ 29, and 16th percentile SPS p(z) $>$ 1; see \citealt{mitsuhashi2026}). 

Overall, we observe that the model-based measurements via photometry (from the outputs of the study \citealt{mitsuhashi2026}) reproduce the 52 spectroscopic measurements albeit with large scatter (see Figure \ref{fig:spec_vs_emp}, see also Figure 2 in \citealt{mitsuhashi2026}). We find that 5 out of 8 spectroscopically confirmed nappers are recovered robustly in photometry-based values of the indicators; napper 61471 does not have an EW(H$\alpha$+[NII]) measurement due to its high redshift, 59873 does not pass our photometric data quality cuts to have a reliable measurement, and 42420's photometric fits do not converge towards a robust measurement for \ewha\ (the 16-84th percentile range classified it as a napper). Critically, no false positives are observed in this analysis, i.e. the sample of photometric napper candidates is likely pure, albeit incomplete, since constraining weak or no emission lines such as H$\alpha$+[NII] with medium-band photometry, and weak flux breaks like Balmer break (at 3650\AA), is a challenging task. Moreover, each photometric estimate of napper indicators for a given galaxy is consistent within approximately 1$\sigma$ of its spectroscopic measurement, as visualized in Figure \ref{fig:spec_vs_emp} by the dotted 1:1 lines. This is therefore a conservative selection, since for a majority of galaxies, median \ewha is systematically overestimated in photometry -- due to photometric redshift uncertainties, photometric filters do not precisely capture $H\alpha$ emission line strengths, and can ``fill in" stronger emission either in between or at the edges of filters sampling $H\alpha$.

\begin{figure*}[htb!]
\centering
\includegraphics[width=1.0\textwidth]{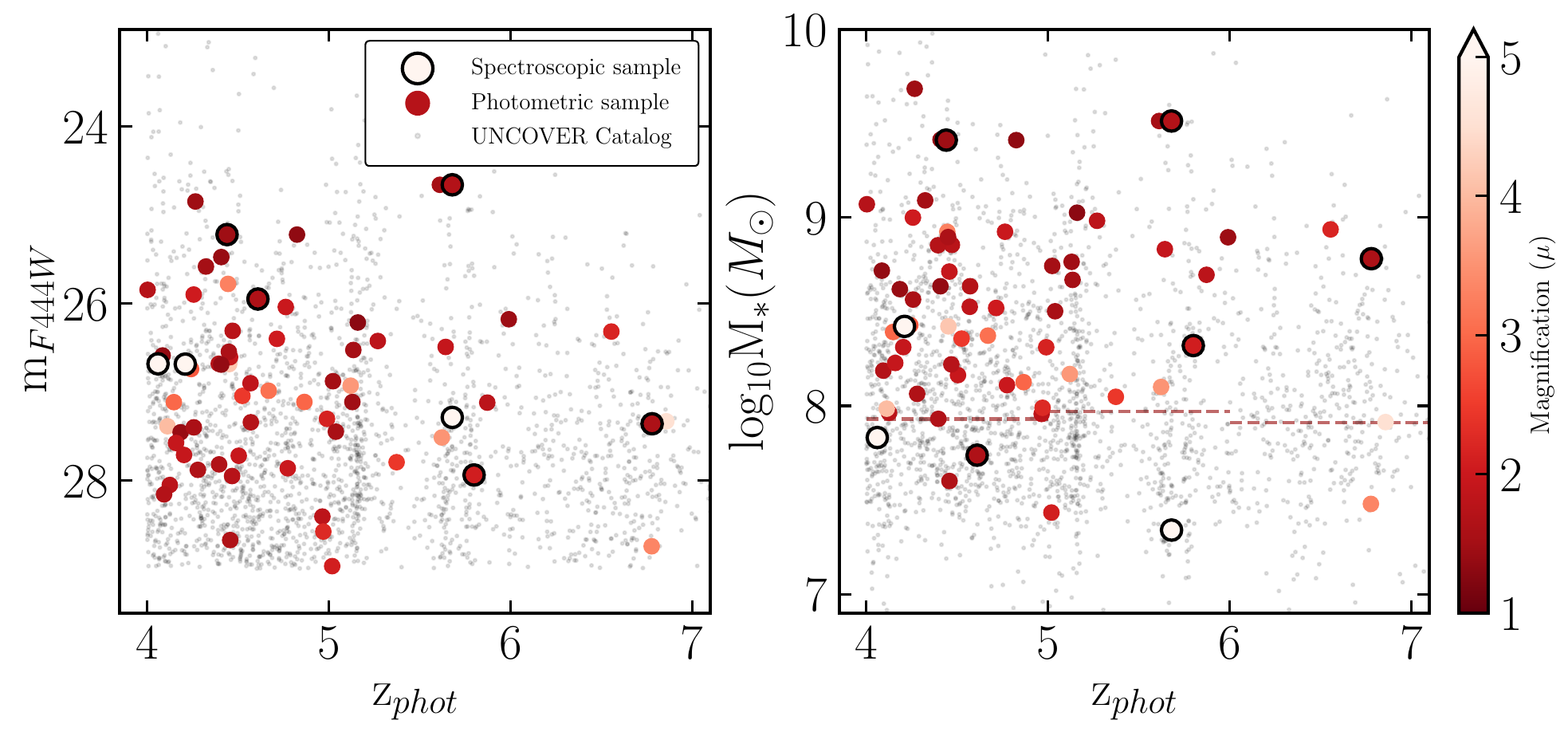}
\caption{Apparent magnitudes (F444W) (left) and stellar mass (right) versus redshift of nappers (colored by magnification) and the parent sample (gray) from the UNCOVER+MegaScience survey. The approximate mass-completeness limit as a function of redshift bin is indicated by the dashed horizontal line. Photometric candidate nappers are indicated by small circles, while spectroscopic sources are highlighted by larger black outlined circles. Nappers have diverse properties, with \logM distributions spanning $\sim$2.5 dex. Most nappers are also found to be well above the photometric and stellar mass completeness limits of the UNCOVER/Megascience survey, with some exceptions being highly magnified objects ($\mu > 5$, in particular the spectroscopic discoveries). The upper limit in redshift for the sample is determined by the JWST/NIRCam F480M sampling of redshifted $H\alpha$ emission.}\label{fig:m444w_zphot}
\end{figure*}

How likely are we to see nappers in a deep photometric survey like UNCOVER+MegaScience (and subsequently, wide field surveys like \JWST MINERVA; \citealt{minerva2025})? We explore various properties that directly impact our ability to detect nappers in UNCOVER MB photometric measurements, and the dependence on the robustness of photometric measurements in key filters -- detection filters, or the filters in which we sample key indicators of nappers. This measurement of $\Delta$(inferred property between photometric and spectroscopic data) is meant to explore the roles of the following factors: p(z) uncertainty (or mismatch between z$_{phot}$ and z$_{spec}$), p(z) -- Balmer Break -- [OII] emission degeneracy, the detection filter, the filter in which Balmer break is nominally sampled, robustness of redshift detection, biased $H_{\alpha}$ estimate in empirical MB measurements because of inconsistent sampling of emission lines (e.g., lines falling in between filter transmission functions), or combination of these factors. We observe that all UNCOVER spectra at $z=4-7$ have a weak dependence on the SNR of the detection filter (F277W + F356W + F444W) as well as the SNR in the filter sampling the blue end of the Balmer break. We discuss this in detail in Appendix \ref{sec:false_positives}.

The outlier objects with the largest $\Delta$(inferred property) -- e.g., objects with DR3 ID 23608, 14411, 8943, 13416, and 27114 -- have uncertain photometric (or spectroscopic) measurements due to a host of reasons, e.g., mismatch between 16-84 percentile p(z) and spec-z ($|z_{spec} - z_{phot}| > 0.1$), mismatch between phot and model (likely due to redshift+filter sampling of OIII or Halpha), the p(z)-OII-Balmer Break degeneracy not being able to solve for dusty SED solutions, and a combination thereof.  

We further explore this by conducting the reverse experiment -- how many photometric napper candidates have been spectrscopicall observed in the DJA, or as part of the UNCOVER/MSA \citep{price2025} and ALT Surveys \citep{alt2024} (in addition to the 5 systems discussed earlier in this section)? A detection in ALT NIRCam Grism or UNCOVER NIRSpec/MSA PRISM/M-grating observations is likely an emission line detection, and hence an indicator that the underlying spectrum is that of a non-napping galaxy.

Of the 60 UNCOVER photometric napper candidates, the DJA v4.4 archival spectral dataset contains ten spectra; these are observations with low signal-to-noise ratio (SNR) continuum observations (these are filtered out in our spectroscopic selection) but show no clear O[III]+H$\beta$ or H$\alpha$+N[II] emission, consistent with napper characteristics -- UNCOVER DR3 IDs 11792, 22439, 31272, 56017 and 57254. UNCOVER 22439, 31272 and 56017 have too low SNR to reliably measure continuum around rest-frame H$\alpha$+N[II] emission, which is a key measurement in estimating napping behavior.


Similarly, out of these photometric candidates, ALT detects six nappers. Three of these candidates also have UNCOVER PRISM spectra -- DR3 IDs 57009 and 47184 ([OIII]+$H\beta$ detection), and 31820 (H$\alpha$+[NII] detection). Remarkably, 57009 is a napper (ALT was able to detect the blended weak [OIII]+$H\beta$ detection in this system, while 47184 lies marginally outside the napper threshold in our study -- moreover, $H\alpha$ emission at 31820's redshift of 4.7 is not sampled robustly by any of NIRCam's filters (F356W, F444W and F360M all miss the emission line, or contain the emission line at the edge of a filter transmission). We also note that each of these detections is at the edge of the Grism transmission/response function.

Additionally, UNCOVER 49632, 33208 (with H$\alpha$+[NII]) and 18549 (with [OIII]+$H\beta$) -- are also detected in ALT (two of them containing emission lines at the edge of the transmission function for the Grism as well). It is likely that H$\alpha$+[NII] detections are star-forming galaxies (false positives in our study), but we are unable to rule out 18549 as a napper. Finally, all three of these objects have a p(z)-$z_{spec}$ mismatch of $|\Delta z| = 0.025 \times(1+z)$, substantial enough to bias MB fits and mischaracterize emission line strengths (or lack thereof). Hence, we posit that we detect 2-5 false positives of the ``negative" sample of 60 photometric candidates ($<10\%$).

To summarize, we find that certainty in p(z) measurements, high SNR of filters sampling the blue end of the Balmer break (which is redshift-dependent), and robustness in photometric sampling of emission features like $H\beta$, $H\alpha$ (or lack thereof) reliably reduce the number of false positives, and lead to a pure sample of nappers in photometric surveys (like UNCOVER+MegaScience, where the observations are effectively an $R\sim15$ low-resolution ``spectrum").

\section{A Census of Napping Galaxies} \label{sec:census}

\subsection{Nappers span $\sim$2.5 dex in stellar mass and $\sim$4 magnitudes in brightness}

Figure \ref{fig:m444w_zphot} shows the observed and inferred properties of UNCOVER nappers as a function of redshift. The left panel shows photometric and spectroscopic nappers (in color) compared with a magnitude-limited (clean) sample of UNCOVER galaxies between $z_{phot}=4-6.6$ (where the upper limit in redshift is determined by the JWST/NIRCam F480M sampling of redshifted $H\alpha$ emission). The right panel shows the stellar mass of nappers as a function of redshift, with a redshift-dependent mass-complete limit of log$_{10}$M$_*$ $\sim 8$, in units of solar masses, derived specifically for the UNCOVER photometric survey; we use an empirical estimate via methodology followed in \cite{Muzzin2013} and \cite{Tomczak:14} (while excluding high magnification -- $\mu > 5$ -- objects from this empirical estimate).

We see that napper galaxies span $\sim$4 magnitudes in F444W, and don't show any obvious pattern in their distribution of observed flux. The magnification-corrected stellar masses (inferred via \texttt{Prospector} SED fitting) for nappers span $\sim 2.5$ orders of magnitude, i.e., we see low (log$_{10}$M$_*$ $< 8$), intermediate (log$_{10}$M$_*$ $\sim 8-9.5$) and high-mass (log$_{10}$M$_*$ $> 9.5$) galaxies in a lull-phase of star formation. This mass range is different from the assertions in the first napper/mini-quenched galaxies discovery studies, where both simulations and observations suggested that lower-mass galaxy halos lose cold molecular gas due to environmental factors and temporarily lower their star formation \citep{Strait2023, Looser2023b, dome2024, ciesla2024, Asada2024, trussler2024}. 


This study, then, either indicates an extension of the mass regime of mini-quenched galaxies, or the discovery of a unique channel of a quenched phase in stochastic SFH galaxies at $z>4$; the DJA study of nappers in \cite{covelo2025} found massive systems also, in particular at \logm $> 9$.

\begin{figure*}[htb!]
\centering
\includegraphics[width=1.0\textwidth]{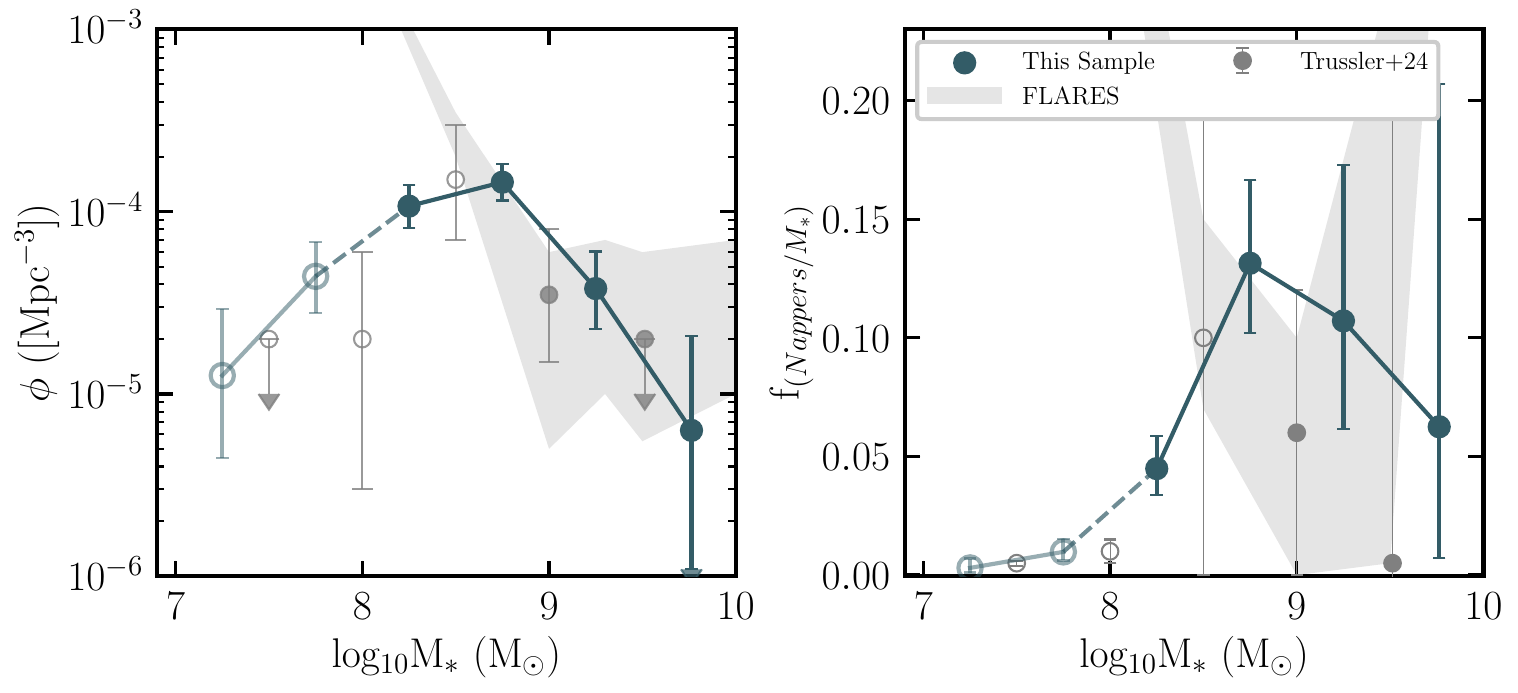}
\caption{The number density (left) and population fraction (right) of napping galaxies as a function of stellar mass (in teal). Error bars only include Poissonian noise, and do not account for cosmic variance. The volumes estimated here are gravitational lensing-corrected. We find similar results to \citet[][black]{trussler2024} at $z=6$, extending to lower masses relatively by $\sim0.5$ dex ($10^{8}M_{\odot}$); we denote mass-complete bins with filled circles. Although we find similar fractions of napping galaxies to the $z=6$ FLARES simulation predictions \citetext{Lovell et al. 2021} at high masses ($\sim10-15\%$ at $10^9M_{\odot}$), we find that the population drops with mass, whereas the fraction rises in simulations. }
\label{fig:mass_function}
\end{figure*}

\subsection{The number density of napping galaxies at $4 < z < 7$}

Photometrically, as we use empirical model-based estimates from broad- and medium-band observations of UNCOVER galaxies, we are likely seeing a pure but nominally incomplete sample of nappers --- weak or absent emission lines are challenging to constrain with this methodology (see discussion in Section \ref{sec:findnappers}). 

Figure \ref{fig:mass_function} shows the estimate of the stellar mass function of nappers from UNCOVER photometric candidates, and the fraction of napper candidates (compared with the overall galaxy population) as a function of stellar mass. Number densities are shown in solid teal circles, where the mass values are greater than the mass-limit of the survey, corresponding to log$_{10}$M$_*$ $\sim 8$, and thus considered complete. Number densities that suffer from sample incompleteness are shown in hollow circles (since we do discover napper candidates in the regime log$_{10}$M$_*$ = 7-8). The uncertainties are assumed to be strictly Poissonian, since we are in a low number regime per mass bin in these estimates. For comparison, we overplot the observations from \cite{trussler2024} at $z\sim6$; the JADES study is mass-complete to log$_{10}$M$_*$ $\sim 9$. This study comprehensively extends the stellar mass function 1 dex lower in stellar mass with significant precision, while remaining consistent with estimates in \cite{trussler2024} at higher masses. We also plot number density and fractional estimates from the FLARES simulations \citep{lovell2021,vijayan2021}; we find our observational stellar mass function and fractional estimates to be consistent, barring uncertainties closest to our mass limit.  In this regime, sample incompleteness sets in for our observational study, while limitations in the FLARES simulations arise when approaching the approximate gas mass resolution regime. We note that closer to our mass limit, we expect the number densities of nappers to be underestimated and/or dependent upon the limiting magnitude in the filter bluewards of the Balmer break; see Section \ref{sec:recovery}). We refrain from referring to this as a lower limit, since the number of objects in this regime are within a few $\sigma$ to that of the Poisson noise in our estimates. The precise bias depends on the break strength in the population, which is influenced by the age of old stellar populations in these systems, and increases (decreases) with the burst amplitude (frequency) of their stochastic star formation histories \citep{mintz2025,mitsuhashi2026,burnham2026}.

We find that 5-15\%\ of galaxies at $z=4-7$ and log$_{10}$M$_*$ = 8-9.5 (within 1$\sigma$ uncertainties) are likely nappers, consistent with both observations in \cite{trussler2024} and the FLARES simulation estimates. Note that FLARES invokes relatively smooth star formation histories and is limited to sampling stellar masses at \logM $>8$, yet still sees 5-35\% of galaxies at \logM $\sim8-9.5$ being dormant. Similarly, \cite{gelli2025} use the SERRA simulations to explore temporary quiescence in $6<z<8$ galaxies, finding that the fraction of napper galaxies (relative to all galaxies) is consistent with zero at \logM$>9$, and 5-20\% (1$\sigma$ uncertainties) -- this is consistent with our assertion at \logM$=9.5-10$, and 1$\sigma$ discrepant at \logM$=9-9.5$.

\begin{figure*}[htb!]
\centering
\includegraphics[width=0.9\textwidth]{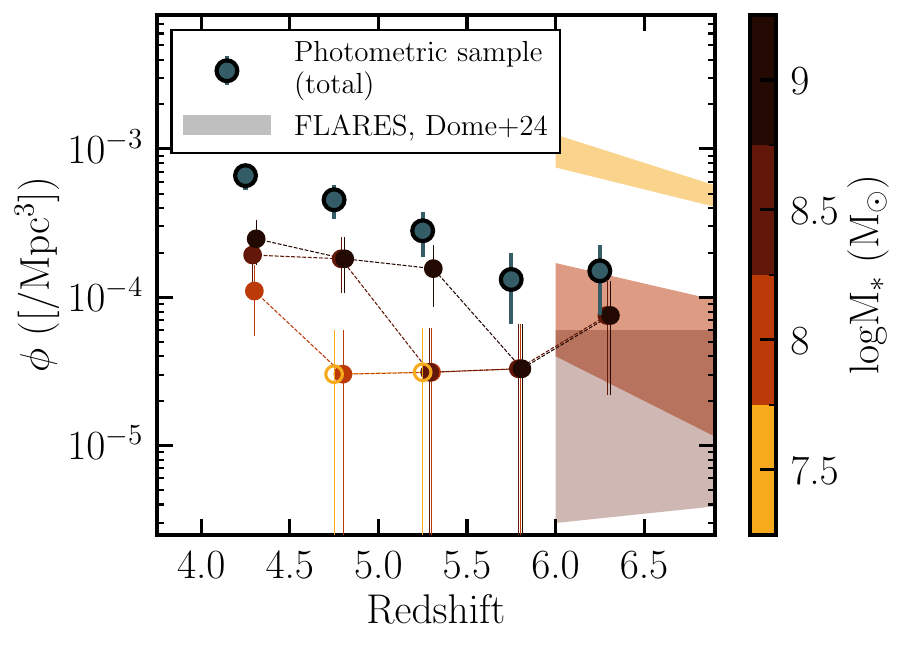}
\caption{
The number density of napping galaxies as a function of redshift. We show napper populations distributed into four mass bins (orange, red, brown and black circles), and total redshift distribution (black squares). Number densities for bins where UNCOVER is mass complete are plotted in filled circles. FLARES \citetext{Lovell et al. 2021} estimates are shown in shaded regions for the corresponding mass bins. Error bars only include Poissonian noise, and do not account for cosmic variance. The volumes estimated here are gravitational lensing-corrected. We find napper densities increases at lower redshifts for galaxies of all masses. We also find marginal evidence for logM $>8$ nappers being more prevalent at low redshifts ($z=4-5$) when compared with low mass nappers. }
\label{fig:mass_function_2}
\end{figure*}

The hallmark feature of this study is our ability to construct a number density of nappers as a function of redshift. In Figure \ref{fig:mass_function_2}, we bin galaxies into four mass bins and 5 redshift bins, which splits our 60 napper candidates adequately, with uncertainties assumed to be Poissonian (neglecting cosmic variance). The total redshift dependent number density is plotted in teal, while mass-dependent curves are shown in their subsequent colors; mass-dependent number densities seen in the FLARES simulation are highlighted with shaded regions (similar to comparisons in \cite{trussler2024}, we use $z\sim6$ and $7$ number densities to construct the shaded regions). 

We find that the number density of nappers decreases with increasing redshift, and there is marginal evidence in the UNCOVER survey for \logM $>8$ nappers being more prevalent at low redshifts ($z=4-5$) when compared with lower mass nappers. The peak of the number density is in the $z=4-4.5$ bin, and the average number density in bins $z=4-5$ is $\Phi (\mathrm{/Mpc^3})= 4-5\times10^{-4}$. For a direct comparison, \cite{yang2026},\cite{2long2024}, \cite{valentino2023}, and \cite{zhang2026} conduct a census of quiescent galaxies at $4<z<5$ to find $\Phi (\mathrm{/Mpc^3})= 1.5-4\times10^{-5}$ (16-84 percentile range); nappers across a wide mass range in this study are at least 5-10$\times$ higher in number density in the redshift range $z=4-5$ relative to quiescent galaxies identified with longer duration suppressed sSFRs.  This redshift regime was not explored previously in post-starburst/napper studies in the literature, and marks an important first comparison point.

\section{Discussion}\label{sec:discussion}

\subsection{Comparing selection methodologies across published studies} \label{sec:comparisons}

The approach to selecting napping/mini-quenched galaxies has been varied across the literature. The primary difference found in napper studies is how SED models are used to isolate these samples. In this study, we find nappers by utilizing our empirical indicators \ewha, Balmer break, and cnUV. The approach seen in the literature, on the other hand, primarily relies on the first two indicators and model-fits to spectroscopic data \citep{covelo2025, Looser2023b}. In these works, only SED fitting-based dust attenuation, mass-weighted age and stellar mass constraints are used to differentiate nappers from massive quiescent galaxies (that are likely to not form stars/stay dormant during their subsequent evolution). Our ethos of inferring napper properties from the data directly requires us to use an empirical indicator like cnUV (rooted in post-starburst SFH shapes, and burstiness indicators seen in studies like \citealt{Looser2023b, munoz2026}, and references therein). Therefore, any difference in our list of nappers compared with model-based estimates in other studies likely results from both the novelty of cnUV as an indicator, as well as the systematic differences between noisy NIRSpec/PRISM spectra vs posterior distributions of SED model fits. For example, \cite{covelo2025} find a multitude of nappers with high values of Balmer break strengths (\texttt{Bagpipes} model-based) which we do not recover (measured via DJA spectra, with uncertainties), including ID 4188 from GO 1433 \citep{hsiao2024} which shows a low SNR spectrum (see Table 1 in \citealt{covelo2025}). Moreover, while our precise selection of \ewha\ and Balmer break strength cuts are similar to \cite{covelo2025} in absolute values (-65 vs -50 \AA\, and 1.35 vs 1.4, respectively), we define these cuts in conjunction with the additional parameter cnUV instead of utilizing SPS models and weeding out dusty galaxies.

Secondly, the SED model templates themselves are different across these studies. For example, \looserb conducted the first spectral search for bursty SFHs and mini-quenching at $z>4$ with the JADES survey, with \code{ppxf} \citep{cappellari2023} spectral fitting and a library of simple stellar population SED templates. They find significant evidence for bursty SFHs at high redshifts, as well as two galaxies in a ``lull" phase -- one already reported in \loosera,  GS-z7-01-QU (at $z=7.3;$ \logM $\sim$ 8.7), and another at $z=4.5;$ \logM $\sim$ 7.8). \cite{Endsley2025} find three MSA spectroscopically-confirmed objects with SFR downturns at $z\sim6$; MSA ID 12065 (UNCOVER DR3 ID 22947) is the closest object to our selection (it does not make our cut). \cite{Endsley2025}'s photometric SED fitting approach using a TcSFH (Truncated Constant Star Formation History) model using BEAGLE \citep{beagle2016} leads to an SFH solution with low sSFR (at the epoch of observation) with a moderate ionization parameter -- our interpretation with a joint-spectrophotometric fit of the object agrees with \cite{Endsley2025} on the epoch of SF burst (30-50 Myr before the epoch of observation), but finds the current sSFR to be too high to be called a napper. \cite{covelo2025} find a majority of the nappers in the Abell 2744 field as well -- MSA IDs 44493, 42420, 26648 and 61471 -- from our study between $z=4-7$ by using Balmer break and \ewha estimators on BAGPIPES models; we incorporate their DJA-published nappers in Figure \ref{fig:nap_parameter_space} to discuss the full population of nappers discovered at these epochs, barring a few systems (as discussed above).

\subsection{Modeling star formation histories at high redshifts with low vs high resolution spectra} 

Since we measure SFRs directly from spectroscopic continuum-based SED fits, it is important to rule out systematics that may be influencing our measurements of the SFHs. 

First, modeling star formation histories is itself a challenging task, as recent studies have demonstrated aptly \citep{iyer2019, leja:2020}. While robust candidates have been confirmed via \texttt{Bagpipes} \citep{carnall:18} using double power-law models for SFHs \citep{trussler2024, covelo2025}, the systematic biases in parametric SFH models is likely to come into play \citep{carnall2019_sfh, Leja2019_sfh}, when comparing with non-parametric fixed-bin models in napper studies \citep{Looser2023a}. 

Second, while our precise inference of SFH timescales is conducted via the most flexible form of SFH models -- specifically calibrated for post-starburst galaxies, the ``continuity-PSB" SFH model within \texttt{Prospector} -- our population-level assertions on the diversity of napper properties, the stellar mass function, and fractional number densities are conducted using empirical model-based measurements; this is a direct extension of our philosophy of conducting measurements as close to the data as possible (see Section \ref{sec:findnappers} for more details).

Moreover, just as \loosera, we observe relatively shallow UV slopes and small H$\delta$ absorption in all our galaxies --- it is unlikely that only a high escape-fraction of ionizing photons is causing low nebular emission \citep{faisstmorishita2024}; we are likely not observing emission because of post-burst SFHs (or at the very least, a combination of high escape-fraction and quenching activity within these galaxies). Despite these assertions, a comprehensive rest-frame UV-to-optical coverage with spectrographs (e.g., NIRSpec M and H gratings) can enable further clarification on this question.

\subsection{Burstiness timescales in the era of JWST}

How bursty are galaxies in the early Universe? How long do these bursts of star formation last in these systems? What causes a galaxy to go through episodes of burst? The answers to these questions are fundamentally connected to the physical processes driving napping behavior \citep{Looser2023b, dome2024, trussler2024}. JWST's spectroscopic data (including the spectra shown in this work) show that while PRISM resolution grants us access to the rise and fall of the most recent bursty SF episode, it is unlikely that a spectrum of a single system (or even a handful of galaxies), however high-resolution it may be, has the constraining power to measure stochasticity in a given SFH. How do we constrain burstiness timescales in galaxies then?  

A first step forward is observations on a statistical sample of high-redshift galaxies, with \JWST/NIRCam MB photometry offering the most obvious path. For example, \cite{mitsuhashi2026} investigate in detail galaxy star formation histories at $z=3-9$ as a function of stellar mass and optical flux, and find no significant evolution in burstiness of SF at $z\sim3-7$; their measurements of empirical line ratios like $H\alpha + N[II]$-to-UV from MB photometry in UNCOVER galaxies do not depend significantly on stellar mass (especially in the range \logm $\sim8-9.5$) or rest-frame optical flux. \citet{mitsuhashi2026} also compare their population-level measurements to FSPS population models; observations favor rising, long-duration and large-amplitude bursts and are agnostic to redshift and mass ranges. 

On the modeling front, the true constraints on stochastic phases of SF would emerge from directly analyzing fluctuations in galaxy SFHs. Firstly, stochastic modeling of SFHs has recently been introduced as a framework in simulations \citep[e.g.,][]{caplar2019}, where star formation regulator models of galaxy formation encode small-scale processes related to the life-cycle of giant molecular clouds (GMCs) within galaxies, to processes operating on larger scales, such as the accretion of gas onto galaxies, and major mergers that cause mass assembly and SF activity disruption \citep{tacchella2020,iyer2024}. A promising methodology to observe these fluctuations in simulations is a power spectral density (PSD) analysis. For example, using \code{densebasis}, an SPS modeling code constructing SFHs via Gaussian Processes (GPs), theoretical SFHs with added stochasticity, and toy models of high-redshift galaxy evolution, \cite{iyer2024} demonstrate that PSDs in high-redshift galaxies can be evaluated with some robustness. 

Secondly, using ML techniques like simulation-based inference (SBI) is a step forward in constraining burst timescales and amplitudes at a population level \citep{burnham2026} --- by adopting a PSD formalism spanning 1 Myr-10 Gyr and a simulated sample of 500 $z\sim 4$ NIRSpec-observed galaxies, \cite{burnham2026} recover small scale ($<100$ Myr) fluctuations (seen in FIRE-2-like and Illustris-like feedback-based galaxies); according to \cite{burnham2026}, with a uniformly selected sample of galaxies this methodology is expected to work for an order of $\sim100$ systems. Galaxies also seem to undergo high amplitude and frequent bursts, observable features of long-term fluctuations are obscured, underestimating the level of burstiness at $<100$ Myr timescales; a detailed explanation of systematic biases in such inferences is described in Section 6 of \cite{burnham2026}. This work also lays out a prescription for how the growing repository of NIRSpec/PRISM spectra, which sample key spectral features constraining the parameters of interest, combined with a physical SPS stochastic model can be implemented within an SBI framework to characterize burstiness, and even quenching timescales of a population of nappers.

In short, with upcoming uniform and deep sampling of large-area surveys, beyond individual lensing cluster fields, we expect to measure a statistically significant sample of spectroscopically confirmed nappers in the near future (e.g., JWST Cycle 5 GO programs like MINERVA, SPAM, the medium-band survey followup of CANUCS fields), without which building a general model of stochastic SFHs in high-redshift galaxies is challenging. 

\subsection{Mini-quenching and galaxy-galaxy interactions}

In the literature, there is a well established connection between galaxy interactions, minor mergers, starburst activity, and subsequent rapid quenching; mergers are expected to trigger both enhanced SF \citep{barnes1998, lin2007} and AGN activity \citep{Hopkins2006,Hopkins2009,Hickox2009,Moreno2015}. This is a natural expectation in a hierarchical merger framework of structure formation, empirically well-inferred for massive galaxies from a wide variety of probes, including comparing size/radial profiles of galaxies at high redshifts to local systems \citep{bezanson2009b,VanDokkum2010,hill2017,whitaker2026}. SDSS observations of galaxies (global, or not spatially resolved) at $z<1$ with SDSS find an excess of mergers in PSB fields \citep{Li2023}, and JWST photometry-based analyses \citep{Suess2022c,Asada2024} find an excess of blue galaxies (labeled ``buddies") in the immediate 2D projected neighborhood ($<$ 2-4") of quenched systems at $0.5 < z < 6$. Most similar analyses expect minor merger mass ratio of $\sim$ 1:10 to 1:100 in PSB environments \citep{French2021,Li2023,Wilkinson2021,Suess2022a}. We observe a subset of our sample (e.g., nappers 57009 and 64792) to have either disturbed morphologies or contain nearby blue galaxies (at similar redshift, and smaller in angular size). This trend continues when we examine the field of JADES-GS-z7-01-QU \citep{Looser2023a} at $z=7.3$, where JWST/NIRCam F115W, F210M and F410M imaging clearly separate a dominant red component from a blue clump $\sim$0.2" ($\sim$1 kpc) west (see \cite{faisstmorishita2024} for a study of variable dust attenuation in the two components). We will explore quantitatively the morphology and environment of these systems in a future publication.


\section{Conclusion} \label{sec:summary}

In this work, we present a census of ``napping'' (mini-quenched / post-starburst-like) galaxies in the UNCOVER/MegaScience observations of the Abell 2744 lensing field, and quantify how common these galaxies are at $z=4-7$. We find 8 spectroscopically confirmed nappers and 60 photometric candidates at $z\simeq4$--7 (with the photometric sample spanning $z=4$--6.6). In this study, we select nappers by their weak or absent nebular emission, demonstrated by low \ewha, a weak Balmer/4000\AA\ break, and relatively blue UV continua (cnUV) similar to young post-starburst galaxies. These criteria allow us to distinguish nappers from dusty or older quenched systems. We find that medium-band photometry can identify incomplete but pure samples of napping galaxies: 5/8 spectroscopic nappers are recovered with medium-band photometry alone, but no false positives are found. We use \texttt{Prospector} to model the SFHs of both our spectroscopic and photometric napper samples with flexible non-parametric models and find that they have bursty, rapidly varying SFHs; nappers are strong outliers in the burstiness metric $\log(\mathrm{SFR}_{10}/\mathrm{SFR}_{100})$ and show trajectories that would be consistent with moving off the SFMS toward quenching if the fuel supply is not replenished.


However, ultimately spectroscopic datasets hold more sensitive diagnostic power to accurately recover stellar population properties. While we show that the {\it selection} of nappers is accurate and pure with medium-band photometry, fitting our spectroscopic sample with medium-band photometry alone reveals that the {\it detailed characterization} can be tricky. The SFHs derived from medium-band photometry capture the same bursts and quenching episodes as the full spectroscopic fit roughly half of the time (4/8 galaxies), but fail to match the quenching epoch recovered from the spectroscopic fit (estimated to be the more accurate SFH). Medium bands are particularly reliable at redshifts where the Balmer break and H$\alpha$ emission line are well-sampled. This implies that using narrow redshift cuts to select samples of nappers may lead to an increase in both sample completeness and SFH model accuracy. Broad-band photometry alone is fundamentally unreliable for this population: not only can broad-band photometry not accurately select nappers, but SPS modeling is unable to constrain sudden quenching episodes in the last 10-20 Myr of a galaxy's SFH.  

While napping galaxies are relatively rare, we find that they constitute a substantial fraction ($\sim5$--15\%) of the population of $\log_{10}(M_*/M_\odot)=8$--10 galaxies at $z=4$--7. Our study suggests that their number density declines toward higher redshift; at $z=4$--5 we measure an average $\Phi\simeq(4$--$5)\times10^{-4}\,\mathrm{Mpc}^{-3}$, which is $\gtrsim5$--10$\times$ higher than published massive-quiescent-galaxy number densities \citep{2long2024} at similar redshifts ($\Phi\simeq(1.5$--$4)\times10^{-5}\,\mathrm{Mpc}^{-3}$; 16--84 percentile range). Our estimate of number densities remain complete to lower stellar masses than previous work, and are consistent with bursty star formation in simulations (FLARES; \citealt{lovell2021,vijayan2021}), as well as other photometric studies \citep{cole2023,trussler2024}.

One of the driving goals of understanding galaxy formation and evolution is characterizing how galaxies form stars and build their stellar mass across cosmic time. Burstiness is an important part of the star formation process, especially at high redshift; however, the exact timescales of this process are not fully understood. Quantifying the stochasticity of star-formation is intractable using archaeological approaches because the youngest generations of stars will vastly outshine their predecessors. This means that we must simultaneously study the full population -- including both the high and low points of bursty star formation -- in order to understand their full life cycle. The high points (starbursts, extreme emission-line galaxies) can be easily identified and studied from high and low-resolution spectroscopy or even from photometry. Our study shows that detecting the lulls in star formation (nappers) does not necessarily require spectroscopy: pure samples of nappers can largely be identified from a combination of broad+medium-band photometry, and, with some care, their SFHs can be accurately recovered. This implies that combining existing and upcoming wide-area medium-band surveys, such as MINERVA \citep{minerva2025}, SPAM (PID 8559), JEMS/JOF (PID 1963; \citealt{jems2023}), and CANUCS/Technicolor (PID 3362; \citealt{technicolor2026}) will be up to the task of fully characterizing the episodic nature of star formation in the early Universe.

\acknowledgements 

GK acknowledges that a significant part of our work in North America is done on stolen land, and we support the efforts of the Land Back movement and true stewardship to those same peoples whose land is occupied. GK also notes that we do not use the full name of \jwst\ due to the person after whom this telescope is named and their role as NASA administrator during the ``Lavender Scare'', as per the $\#RenameJWST$ protest movement.

GK would like to thank the Baum Grant and Fellowship at the University of Washington for support during this work, as well as the ALMA Ambassador Program (administered by NAASC and NRAO). GK is also grateful to the International Space Science Institute (ISSI), Bern, for their hospitality, financial support and collaboration during the time of writing this manuscript. GK would also like to thank the DiRAC Institute in the Department of Astronomy at the University of Washington. The DiRAC Institute is supported through generous gifts from the
Charles and Lisa Simonyi Fund for Arts and Sciences, Janet and Lloyd Frink, and the Washington Research Foundation. 

RB acknowledges support from the Research Corporation for Scientific Advancement (RCSA) Cottrell Scholar Award ID No: 27587. Cloud-based data processing and file storage for this work is provided by the AWS Cloud Credits for Research program. TBM was supported by a CIERA Fellowship. DJS and JRW acknowledge that support for this work was provided by The Brinson Foundation through a Brinson Prize Fellowship grant. ORC acknowledges support from National Science Foundation Astronomy \& Astrophysics Postdoctoral Fellowship Award No. 2503202. AdG acknowledges support from a Clay Fellowship awarded by the Smithsonian Astrophysical Observatory. AdG acknowledges funding by the European Union (ERC, FIRST-GIANTS, 101221926). AdG is supported by the Lise Meitner Excellence program of the Max Planck Society.

This research was supported in part by the University of Pittsburgh Center for Research Computing, RRID:SCR\_022735, through the resources provided. Specifically, this work used the H2P/MPI cluster, which is supported by NSF award number OAC-2117681. Some data products presented herein were retrieved from the Dawn JWST Archive (DJA); DJA is an initiative of the Cosmic Dawn Center (DAWN), which is funded by the Danish National Research Foundation under grant DNRF140. 

This work is based in part on observations made with the NASA/ESA/CSA \emph{JWST}. The data were obtained from the Mikulski Archive for Space Telescopes at the Space Telescope Science Institute, which is operated by the Association of Universities for Research in Astronomy, Inc., under NASA contract NAS 5-03127 for \JWST. These observations are associated with \JWST Cycle 1 GO programs \#2561, \#4111, JWST-ERS-1324, JWST-DD-2756. The specific observations and catalogs analyzed can be accessed via \dataset[https://doi.org/10.17909/1ms3-sr76]{https://doi.org/10.17909/1ms3-sr76}, \dataset[https://zenodo.org/records/11059273]{https://zenodo.org/records/11059273}, and \dataset[https://doi.org/10.17909/8k5c-xr27]{https://doi.org/10.17909/8k5c-xr27}.

Support for US investigators in program JWST-GO-2561 was provided by NASA through a grant from the Space Telescope Science Institute, which is operated by the Association of Universities for Research in Astronomy, Inc., under NASA contract NAS5-03127. This research is based on observations made with the NASA/ESA Hubble Space Telescope obtained from the Space Telescope Science Institute, which is operated by the Association of Universities for Research in Astronomy, Inc., under NASA contract NAS 5–26555. These observations are associated with programs HST-GO-11689, HST-GO-13386, HST-GO/DD-13495, HST-GO-13389, HST-GO-15117, and HST-GO/DD-17231.

This publication used the large language model Claude Sonnet 4.5 specifically for debugging code which was written for plotting figures and formatting LaTex tables presented here. 


\vspace{5mm}
\facilities{\JWST{}(NIRCam, NIRSpec, and NIRISS), \HST{}(ACS and WFC3)}

\software{\code{Python 3.6 --- Prospector \citep{johnson21}, python-FSPS \citep{Conroy2009,Conroy2010}, SEDpy, Matplotlib \citep{Hunter:2007}, Numpy \citep{numpy2011,numpy2020}, Scipy \citep{scipy2020}, Astropy \citep{astropy:2013,astropy:2018,2022ApJ...935..167A}, Jupyter, IPython} Notebooks, Grizli \citep{brammer:2021}, DrizzlePac \citep{Gonzaga:12}}

\newpage

\bibliography{all}

\appendix

\section{SED Fits for spectroscopically confirmed UNCOVER Nappers} \label{sec:sed_fits_appendix}

In Figures \ref{fig:sedfit_2} and \ref{fig:sedfit_3}, we show the SPS modeling results for the remaining five spectroscopically confirmed UNCOVER nappers -- MSA IDs 29094, 31249, 64792, 42420 and 59873. Similar to the results demonstrated in Figure \ref{fig:burstiness_2}, our best fits for \code{Prospector} spectrophotometric as well as photometry-only modeling show that medium bands occasionally strongly/marginally agree with spectrophotometric SFH results. In most cases broad-band only fits either do not recover the burst and quenching episode, or miss a proper characterization of the photometric redshift of the object itself.

\section{Stellar population properties of photometric nappers}

Table \ref{tab:nap_emp_sps_part1} and \ref{tab:nap_emp_sps_part2} show the physical properties of all photometric nappers identified in Abell 2744.

\begin{deluxetable*}{cc|ccc|ccc}
\tablecaption{Photometric Post-Starburst Candidates from UNCOVER with SPS Parameters (Part 1)}
\label{tab:nap_emp_sps_part1}
\tablewidth{0pt}
\tablehead{
\colhead{ID} & 
\colhead{$z_{\rm phot}$} & 
\colhead{Balmer Break} & 
\colhead{EW(H$\alpha$+[NII])} & 
\colhead{cnUV} & 
\colhead{$\log(M_\star/M_\odot)$} & 
\colhead{$\tau_V$} & 
\colhead{Age$_{\rm MW}$} \\
\colhead{} & 
\colhead{} & 
\colhead{} & 
\colhead{[\AA]} & 
\colhead{} & 
\colhead{} & 
\colhead{} & 
\colhead{[Gyr]}
}
\startdata
19111 & 6.56 & $1.21_{-0.14}^{+0.14}$ & $-100_{-72}^{+141}$ & $1.09_{-0.08}^{+0.10}$ & $7.91_{-0.39}^{+0.14}$ & $0.05_{-0.03}^{+0.05}$ & $0.09_{-0.04}^{+0.04}$ \\
17540 & 6.51 & $1.44_{-0.10}^{+0.12}$ & $-65_{-32}^{+65}$ & $1.01_{-0.05}^{+0.05}$ & $8.94_{-0.16}^{+0.12}$ & $0.11_{-0.08}^{+0.20}$ & $0.17_{-0.04}^{+0.05}$ \\
16939 & 6.07 & $1.35_{-0.33}^{+0.37}$ & $-234_{-169}^{+325}$ & $0.54_{-0.12}^{+0.16}$ & $7.48_{-0.28}^{+0.30}$ & $0.11_{-0.07}^{+0.12}$ & $0.17_{-0.11}^{+0.10}$ \\
57535 & 6.05 & $1.52_{-0.13}^{+0.19}$ & $-100_{-51}^{+58}$ & $0.48_{-0.03}^{+0.04}$ & $8.90_{-0.05}^{+0.06}$ & $0.07_{-0.04}^{+0.05}$ & $0.11_{-0.04}^{+0.07}$ \\
43561 & 5.85 & $1.44_{-0.11}^{+0.12}$ & $-71_{-46}^{+71}$ & $0.41_{-0.03}^{+0.03}$ & $8.70_{-0.12}^{+0.11}$ & $0.27_{-0.13}^{+0.16}$ & $0.23_{-0.06}^{+0.06}$ \\
35353 & 5.66 & $1.52_{-0.21}^{+0.24}$ & $-161_{-103}^{+160}$ & $0.69_{-0.07}^{+0.07}$ & $8.10_{-0.20}^{+0.20}$ & $0.13_{-0.09}^{+0.44}$ & $0.19_{-0.08}^{+0.07}$ \\
47184 & 5.66 & $1.52_{-0.17}^{+0.21}$ & $-109_{-61}^{+111}$ & $0.38_{-0.04}^{+0.04}$ & $8.83_{-0.10}^{+0.09}$ & $0.17_{-0.06}^{+0.08}$ & $0.21_{-0.05}^{+0.08}$ \\
57009 & 5.62 & $1.61_{-0.06}^{+0.06}$ & $-69_{-16}^{+17}$ & $0.58_{-0.01}^{+0.01}$ & $9.51_{-0.07}^{+0.11}$ & $0.03_{-0.02}^{+0.05}$ & $0.18_{-0.07}^{+0.13}$ \\
26648 & 5.62 & $1.50_{-0.05}^{+0.05}$ & $-40_{-26}^{+37}$ & $0.87_{-0.02}^{+0.03}$ & $7.34_{-0.06}^{+0.11}$ & $0.04_{-0.03}^{+0.08}$ & $0.11_{-0.04}^{+0.07}$ \\
26647 & 5.53 & $1.40_{-0.10}^{+0.09}$ & $-59_{-36}^{+97}$ & $1.11_{-0.07}^{+0.08}$ & $7.22_{-0.07}^{+0.07}$ & $0.03_{-0.02}^{+0.13}$ & $0.19_{-0.06}^{+0.05}$ \\
50886 & 5.30 & $1.77_{-0.17}^{+0.18}$ & $-38_{-23}^{+37}$ & $0.60_{-0.04}^{+0.04}$ & $8.98_{-0.10}^{+0.08}$ & $0.10_{-0.07}^{+0.14}$ & $0.28_{-0.06}^{+0.06}$ \\
20442 & 5.19 & $1.72_{-0.23}^{+0.25}$ & $-76_{-53}^{+113}$ & $0.63_{-0.06}^{+0.07}$ & $8.05_{-0.11}^{+0.16}$ & $0.04_{-0.03}^{+0.18}$ & $0.19_{-0.06}^{+0.07}$ \\
67055 & 5.19 & $1.75_{-0.13}^{+0.15}$ & $-105_{-44}^{+48}$ & $0.34_{-0.02}^{+0.02}$ & $9.03_{-0.08}^{+0.07}$ & $0.14_{-0.06}^{+0.14}$ & $0.25_{-0.06}^{+0.07}$ \\
17605 & 5.19 & $1.55_{-0.06}^{+0.08}$ & $-36_{-20}^{+36}$ & $0.73_{-0.02}^{+0.02}$ & $8.76_{-0.10}^{+0.08}$ & $0.09_{-0.07}^{+0.18}$ & $0.24_{-0.06}^{+0.07}$ \\
57414 & 5.16 & $1.94_{-0.13}^{+0.14}$ & $-93_{-39}^{+43}$ & $0.62_{-0.02}^{+0.03}$ & $8.67_{-0.06}^{+0.14}$ & $0.02_{-0.01}^{+0.09}$ & $0.18_{-0.07}^{+0.08}$ \\
29232 & 5.12 & $1.60_{-0.12}^{+0.11}$ & $-85_{-45}^{+50}$ & $0.63_{-0.02}^{+0.03}$ & $8.17_{-0.10}^{+0.10}$ & $0.03_{-0.02}^{+0.12}$ & $0.21_{-0.10}^{+0.08}$ \\
21812 & 5.07 & $1.90_{-0.19}^{+0.17}$ & $-83_{-49}^{+63}$ & $0.36_{-0.02}^{+0.02}$ & $8.74_{-0.08}^{+0.09}$ & $0.11_{-0.07}^{+0.19}$ & $0.32_{-0.06}^{+0.06}$ \\
16477 & 5.06 & $1.70_{-0.41}^{+0.42}$ & $-155_{-103}^{+174}$ & $0.71_{-0.10}^{+0.11}$ & $7.96_{-0.20}^{+0.15}$ & $0.12_{-0.09}^{+0.26}$ & $0.20_{-0.07}^{+0.10}$ \\
21989 & 5.01 & $1.77_{-0.17}^{+0.23}$ & $-86_{-60}^{+103}$ & $0.69_{-0.04}^{+0.05}$ & $8.50_{-0.12}^{+0.11}$ & $0.06_{-0.05}^{+0.14}$ & $0.20_{-0.07}^{+0.08}$ \\
23352 & 4.98 & $1.04_{-0.39}^{+0.39}$ & $-191_{-134}^{+246}$ & $1.39_{-0.18}^{+0.20}$ & $7.43_{-0.19}^{+0.15}$ & $0.01_{-0.01}^{+0.02}$ & $0.17_{-0.08}^{+0.10}$ \\
15864 & 4.97 & $1.96_{-0.32}^{+0.30}$ & $-129_{-89}^{+144}$ & $0.69_{-0.05}^{+0.07}$ & $8.31_{-0.16}^{+0.16}$ & $0.06_{-0.05}^{+0.11}$ & $0.23_{-0.08}^{+0.14}$ \\
22439 & 4.92 & $1.57_{-0.30}^{+0.38}$ & $-147_{-99}^{+165}$ & $0.94_{-0.12}^{+0.13}$ & $7.99_{-0.21}^{+0.13}$ & $0.10_{-0.08}^{+0.20}$ & $0.24_{-0.09}^{+0.10}$ \\
32076 & 4.87 & $1.35_{-0.09}^{+0.07}$ & $-21_{-15}^{+25}$ & $0.81_{-0.02}^{+0.02}$ & $8.13_{-0.13}^{+0.26}$ & $0.22_{-0.20}^{+0.37}$ & $0.06_{-0.02}^{+0.06}$ \\
64792 & 4.87 & $1.81_{-0.57}^{+0.51}$ & $-122_{-82}^{+113}$ & $0.40_{-0.04}^{+0.04}$ & $9.41_{-0.12}^{+0.12}$ & $0.14_{-0.10}^{+0.29}$ & $0.34_{-0.10}^{+0.11}$ \\
29537 & 4.87 & $3.35_{-0.69}^{+0.88}$ & $-101_{-73}^{+158}$ & $0.54_{-0.07}^{+0.08}$ & $7.56_{-0.19}^{+0.08}$ & $0.05_{-0.04}^{+0.08}$ & $0.27_{-0.11}^{+0.08}$ \\
27124 & 4.77 & $1.58_{-0.78}^{+1.30}$ & $-177_{-128}^{+229}$ & $0.35_{-0.09}^{+0.10}$ & $7.60_{-0.92}^{+0.35}$ & $0.41_{-0.28}^{+0.51}$ & $0.31_{-0.19}^{+0.64}$ \\
34530 & 4.75 & $2.33_{-0.22}^{+0.20}$ & $-21_{-13}^{+21}$ & $0.59_{-0.02}^{+0.02}$ & $8.92_{-0.08}^{+0.08}$ & $0.05_{-0.03}^{+0.05}$ & $0.26_{-0.07}^{+0.07}$ \\
34531 & 4.71 & $3.59_{-0.62}^{+0.86}$ & $-61_{-35}^{+46}$ & $0.60_{-0.04}^{+0.03}$ & $8.52_{-0.09}^{+0.07}$ & $0.01_{-0.01}^{+0.02}$ & $0.25_{-0.08}^{+0.08}$ \\
31820 & 4.69 & $1.34_{-0.27}^{+0.24}$ & $-117_{-56}^{+70}$ & $0.59_{-0.03}^{+0.05}$ & $8.11_{-0.17}^{+0.12}$ & $0.10_{-0.06}^{+0.12}$ & $0.26_{-0.09}^{+0.10}$ \\
34316 & 4.65 & $1.91_{-0.55}^{+0.48}$ & $-149_{-90}^{+108}$ & $0.37_{-0.06}^{+0.06}$ & $8.37_{-0.30}^{+0.15}$ & $0.20_{-0.11}^{+0.13}$ & $0.34_{-0.12}^{+0.13}$ \\
\enddata
\tablecomments{
Photometric nappers selected from UNCOVER DR3 with 
Balmer Break $> 1.35$, EW(H$\alpha$+[NII]) $> -65$ \AA, and 
continuum-normalized UV flux (cnUV) $> 0.2$. 
Stellar mass ($\log(M_\star/M_\odot)$), dust attenuation influencing stellar populations of all ages ($\tau_V$), and mass-weighted age (Age$_{\rm MW}$) are derived from Bayesian SED fitting using \texttt{Prospector} \citep{wangsps2024}. 
Errors represent 16th and 84th percentile confidence intervals. 
Objects are sorted by decreasing photometric redshift. 
Table continues in Part 2.
}
\end{deluxetable*}

\clearpage

\begin{deluxetable*}{cc|ccc|ccc}
\tablecaption{Photometric Post-Starburst Candidates from UNCOVER with SPS Parameters (Part 2)}
\label{tab:nap_emp_sps_part2}
\tablewidth{0pt}
\tablehead{
\colhead{ID} & 
\colhead{$z_{\rm phot}$} & 
\colhead{Balmer Break} & 
\colhead{EW(H$\alpha$+[NII])} & 
\colhead{cnUV} & 
\colhead{$\log(M_\star/M_\odot)$} & 
\colhead{$\tau_V$} & 
\colhead{Age$_{\rm MW}$} \\
\colhead{} & 
\colhead{} & 
\colhead{} & 
\colhead{[\AA]} & 
\colhead{} & 
\colhead{} & 
\colhead{} & 
\colhead{[Gyr]}
}
\startdata
49533 & 4.58 & $1.95_{-0.37}^{+0.80}$ & $-112_{-64}^{+79}$ & $0.35_{-0.05}^{+0.08}$ & $8.64_{-0.14}^{+0.08}$ & $0.13_{-0.07}^{+0.12}$ & $0.40_{-0.12}^{+0.11}$ \\
47857 & 4.56 & $1.63_{-0.19}^{+0.23}$ & $-100_{-48}^{+51}$ & $0.50_{-0.03}^{+0.04}$ & $8.53_{-0.10}^{+0.11}$ & $0.06_{-0.04}^{+0.10}$ & $0.28_{-0.09}^{+0.12}$ \\
59292 & 4.56 & $1.34_{-0.61}^{+0.83}$ & $-152_{-101}^{+163}$ & $0.34_{-0.05}^{+0.07}$ & $8.64_{-0.28}^{+0.16}$ & $0.27_{-0.17}^{+0.25}$ & $0.32_{-0.13}^{+0.23}$ \\
32565 & 4.55 & $1.60_{-0.15}^{+0.19}$ & $-85_{-39}^{+40}$ & $0.54_{-0.03}^{+0.03}$ & $8.36_{-0.07}^{+0.15}$ & $0.06_{-0.04}^{+0.08}$ & $0.25_{-0.08}^{+0.22}$ \\
27507 & 4.54 & $1.73_{-0.25}^{+0.25}$ & $-45_{-33}^{+82}$ & $0.40_{-0.04}^{+0.07}$ & $7.51_{-0.15}^{+0.24}$ & $0.10_{-0.03}^{+0.05}$ & $0.18_{-0.09}^{+0.15}$ \\
56017 & 4.53 & $1.58_{-0.32}^{+0.81}$ & $-115_{-68}^{+86}$ & $0.37_{-0.07}^{+0.05}$ & $8.16_{-0.14}^{+0.20}$ & $0.15_{-0.08}^{+0.12}$ & $0.34_{-0.13}^{+0.16}$ \\
49632 & 4.50 & $1.25_{-0.09}^{+0.11}$ & $-89_{-26}^{+28}$ & $0.75_{-0.02}^{+0.02}$ & $8.85_{-0.04}^{+0.04}$ & $0.02_{-0.01}^{+0.02}$ & $0.25_{-0.03}^{+0.16}$ \\
48858 & 4.48 & $2.76_{-1.34}^{+1.01}$ & $-118_{-74}^{+116}$ & $0.47_{-0.06}^{+0.05}$ & $8.23_{-0.15}^{+0.17}$ & $0.14_{-0.10}^{+0.17}$ & $0.34_{-0.11}^{+0.12}$ \\
58525 & 4.44 & $1.36_{-0.16}^{+0.18}$ & $-84_{-47}^{+48}$ & $0.76_{-0.03}^{+0.03}$ & $8.90_{-0.10}^{+0.07}$ & $0.10_{-0.07}^{+0.10}$ & $0.45_{-0.10}^{+0.08}$ \\
35716 & 4.44 & $1.86_{-0.35}^{+0.60}$ & $-85_{-48}^{+62}$ & $0.38_{-0.03}^{+0.03}$ & $8.92_{-0.16}^{+0.08}$ & $0.16_{-0.11}^{+0.14}$ & $0.36_{-0.15}^{+0.11}$ \\
31249 & 4.43 & $1.53_{-0.24}^{+0.24}$ & $-91_{-60}^{+99}$ & $0.51_{-0.03}^{+0.04}$ & $8.42_{-0.26}^{+0.08}$ & $0.07_{-0.05}^{+0.09}$ & $0.36_{-0.14}^{+0.06}$ \\
59151 & 4.43 & $1.91_{-0.16}^{+0.15}$ & $-30_{-18}^{+22}$ & $0.39_{-0.01}^{+0.01}$ & $9.41_{-0.10}^{+0.10}$ & $0.16_{-0.12}^{+0.17}$ & $0.45_{-0.08}^{+0.07}$ \\
57254 & 4.40 & $1.70_{-0.28}^{+0.34}$ & $-56_{-38}^{+60}$ & $0.36_{-0.04}^{+0.04}$ & $8.22_{-0.13}^{+0.09}$ & $0.13_{-0.05}^{+0.09}$ & $0.31_{-0.08}^{+0.08}$ \\
22565 & 4.39 & $1.53_{-0.37}^{+0.39}$ & $-128_{-72}^{+88}$ & $0.69_{-0.04}^{+0.05}$ & $7.93_{-0.14}^{+0.10}$ & $0.02_{-0.01}^{+0.03}$ & $0.25_{-0.14}^{+0.14}$ \\
54859 & 4.38 & $2.12_{-0.23}^{+0.24}$ & $-43_{-27}^{+35}$ & $0.44_{-0.02}^{+0.02}$ & $8.85_{-0.06}^{+0.04}$ & $0.02_{-0.01}^{+0.03}$ & $0.43_{-0.11}^{+0.07}$ \\
48857 & 4.35 & $2.86_{-0.68}^{+1.02}$ & $-63_{-38}^{+57}$ & $0.39_{-0.04}^{+0.05}$ & $8.71_{-0.09}^{+0.09}$ & $0.07_{-0.04}^{+0.12}$ & $0.37_{-0.09}^{+0.08}$ \\
31272 & 4.31 & $1.22_{-0.21}^{+0.25}$ & $-155_{-99}^{+124}$ & $0.78_{-0.05}^{+0.05}$ & $8.31_{-0.14}^{+0.09}$ & $0.24_{-0.19}^{+0.30}$ & $0.27_{-0.08}^{+0.08}$ \\
58308 & 4.31 & $1.55_{-0.08}^{+0.09}$ & $-54_{-23}^{+21}$ & $0.28_{-0.01}^{+0.01}$ & $9.68_{-0.72}^{+0.07}$ & $0.38_{-0.10}^{+0.17}$ & $0.42_{-0.11}^{+0.16}$ \\
37843 & 4.27 & $1.31_{-0.42}^{+0.45}$ & $-124_{-80}^{+117}$ & $0.59_{-0.06}^{+0.07}$ & $8.43_{-0.13}^{+0.14}$ & $0.11_{-0.05}^{+0.10}$ & $0.23_{-0.08}^{+0.14}$ \\
53201 & 4.26 & $1.92_{-0.25}^{+0.27}$ & $-82_{-41}^{+48}$ & $0.35_{-0.02}^{+0.03}$ & $8.56_{-0.11}^{+0.09}$ & $0.22_{-0.15}^{+0.11}$ & $0.44_{-0.13}^{+0.12}$ \\
14575 & 4.26 & $1.94_{-0.22}^{+0.23}$ & $-81_{-42}^{+47}$ & $0.51_{-0.02}^{+0.02}$ & $9.00_{-0.12}^{+0.09}$ & $0.05_{-0.04}^{+0.07}$ & $0.34_{-0.08}^{+0.14}$ \\
22922 & 4.24 & $1.39_{-0.07}^{+0.06}$ & $-66_{-21}^{+21}$ & $0.75_{-0.01}^{+0.01}$ & $9.09_{-0.18}^{+0.22}$ & $0.20_{-0.16}^{+0.29}$ & $0.25_{-0.12}^{+0.16}$ \\
18549 & 4.24 & $1.26_{-0.15}^{+0.12}$ & $-99_{-48}^{+60}$ & $1.07_{-0.04}^{+0.05}$ & $8.06_{-0.07}^{+0.08}$ & $0.03_{-0.02}^{+0.05}$ & $0.12_{-0.05}^{+0.08}$ \\
27172 & 4.17 & $3.19_{-0.50}^{+0.76}$ & $-58_{-39}^{+59}$ & $0.54_{-0.04}^{+0.03}$ & $8.39_{-0.14}^{+0.08}$ & $0.07_{-0.05}^{+0.11}$ & $0.47_{-0.13}^{+0.11}$ \\
33208 & 4.14 & $2.37_{-0.54}^{+0.81}$ & $-143_{-81}^{+117}$ & $0.29_{-0.07}^{+0.05}$ & $8.19_{-0.16}^{+0.09}$ & $0.13_{-0.09}^{+0.22}$ & $0.52_{-0.15}^{+0.26}$ \\
23881 & 4.13 & $2.93_{-0.60}^{+0.82}$ & $-103_{-55}^{+75}$ & $0.43_{-0.03}^{+0.03}$ & $7.99_{-0.08}^{+0.07}$ & $0.05_{-0.04}^{+0.09}$ & $0.44_{-0.11}^{+0.11}$ \\
11792 & 4.12 & $1.66_{-0.33}^{+0.38}$ & $-99_{-68}^{+98}$ & $0.67_{-0.05}^{+0.06}$ & $7.96_{-0.15}^{+0.17}$ & $0.11_{-0.07}^{+0.09}$ & $0.22_{-0.09}^{+0.15}$ \\
63519 & 4.11 & $1.71_{-0.13}^{+0.19}$ & $-104_{-41}^{+40}$ & $0.67_{-0.03}^{+0.03}$ & $8.72_{-0.09}^{+0.13}$ & $0.04_{-0.03}^{+0.05}$ & $0.29_{-0.11}^{+0.21}$ \\
40511 & 4.06 & $1.54_{-0.15}^{+0.16}$ & $-100_{-43}^{+48}$ & $0.43_{-0.03}^{+0.03}$ & $8.62_{-0.05}^{+0.04}$ & $0.11_{-0.04}^{+0.10}$ & $0.37_{-0.09}^{+0.09}$ \\
11593 & 4.03 & $1.63_{-0.17}^{+0.20}$ & $-67_{-42}^{+51}$ & $0.39_{-0.02}^{+0.03}$ & $9.07_{-0.09}^{+0.10}$ & $0.15_{-0.08}^{+0.10}$ & $0.36_{-0.09}^{+0.43}$ \\
\enddata
\tablecomments{
Continuation of Table \ref{tab:nap_emp_sps_part1}. 
Photometric redshifts and SPS parameters from \texttt{Prospector} SED fitting. 
Errors represent 16th and 84th percentile confidence intervals. 
$\tau_V$ is the V-band optical depth from the Calzetti+00 dust attenuation law. 
}
\end{deluxetable*}

\section{Detectability of Nappers in Photometric MB Measurements -- False Positives and the Impact of Balmer Break Detection} \label{sec:false_positives}

Here, we explore various properties that directly impact our ability to detect nappers in UNCOVER MB photometric measurements, and the dependence of this detectability (and any false positives) on the detection filter, the filter in which Balmer break is nominally sampled, as well as the robustness of redshift detection. We attemp to answer the following questions:

\begin{enumerate}
\item Does the p(z) -- Balmer Break -- [OII] emission degeneracy play a role in MBs not recognizing nappers?
\item Are some rest-frame UV photometry or spectra (or continuum around $H_{\alpha}$ too noisy to reliably estimate cnUV, a key indicator in our study?
\item For specific objects, is the $H_{\alpha}$ estimate in empirical MB measurements biased because of inconsistent sampling of emission lines (e.g., lines falling in between filter transmission functions)? 
\end{enumerate}

\begin{figure*}[htb!]
\centering
\includegraphics[width=1.0\textwidth]{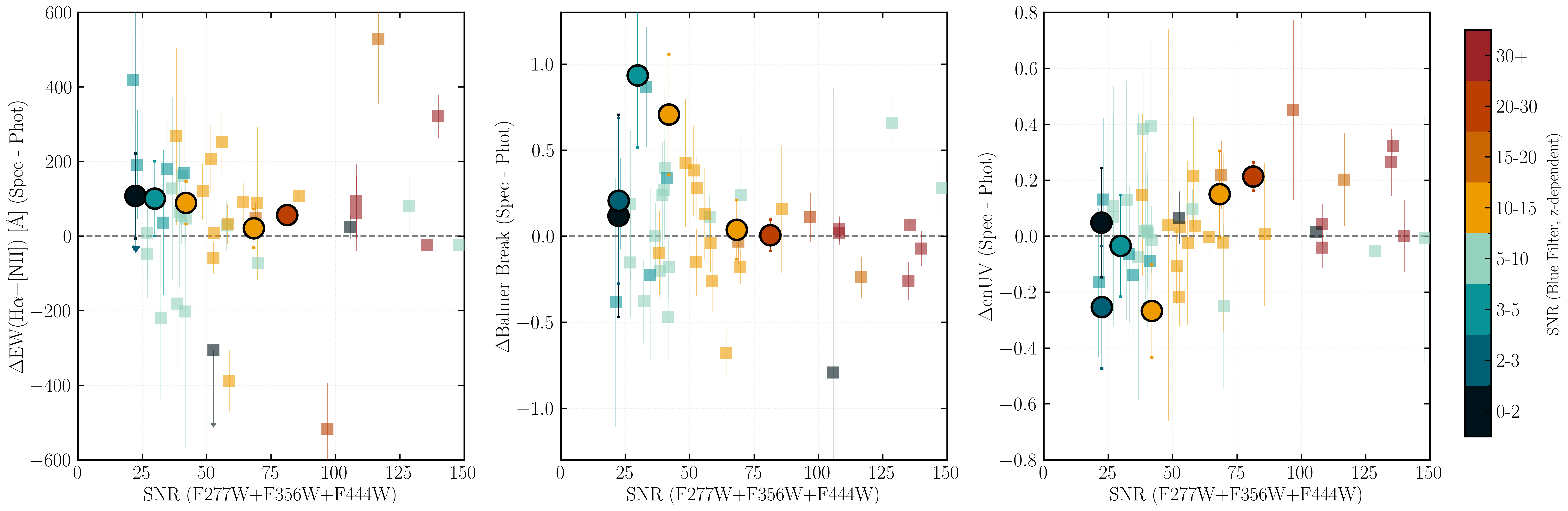}
\includegraphics[width=1.0\textwidth]{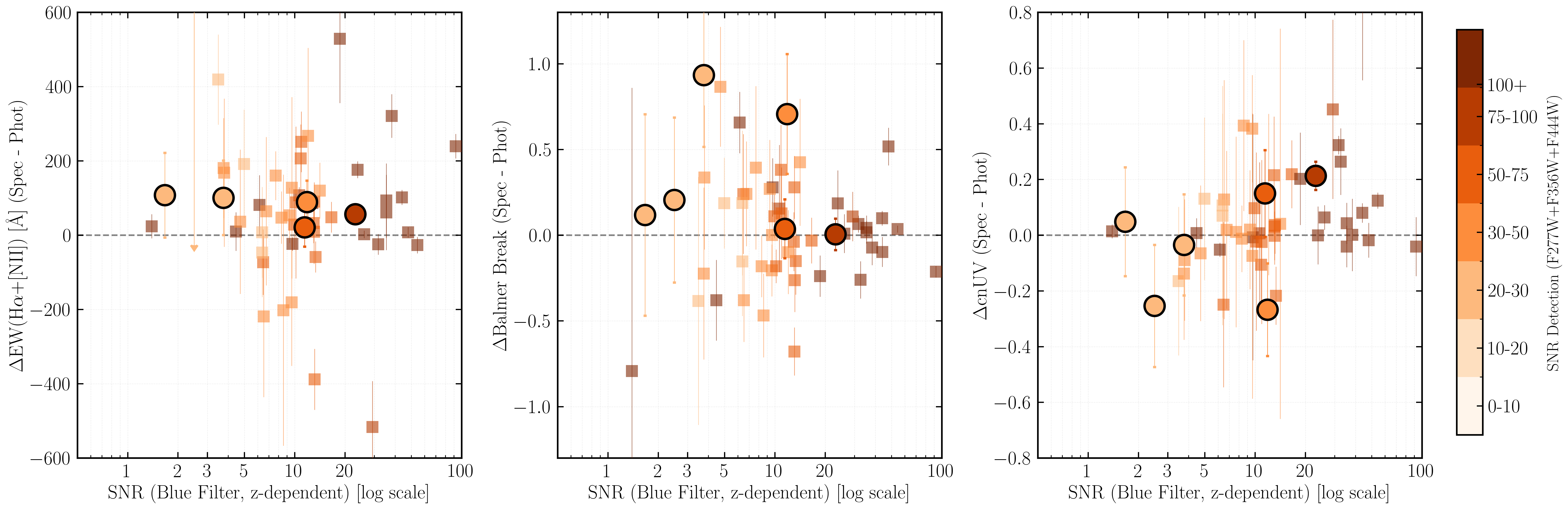}

\caption{(Top) $\Delta$(inferred property) of nappers as a function of SNR (detection), where the color signifies the redshift-dependent filter from MEGASCIENCE which is sampling the blue end of the "Balmer Break" estimate -- a large part of napper non-detectability is due to weak rest-frame UV flux densities and low (SNR) estimates of the Balmer Break. (Bottom) $\Delta$(inferred property) of nappers as a function of SNR (blue Balmer Break filter). The color signifies the SNR(detection). }
\label{fig:delta_1}
\end{figure*}

\begin{figure*}[htb!]
\centering
\includegraphics[width=1.0\textwidth]{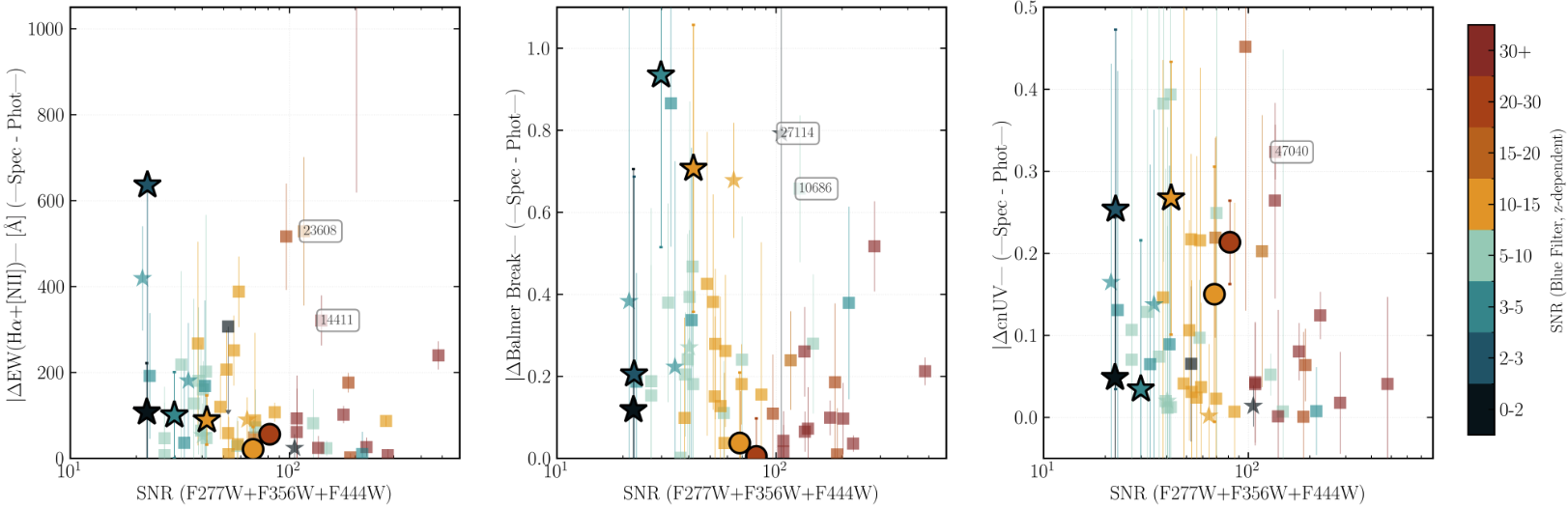}

\caption{ Same as the top subplot in Figure \ref{fig:delta_1}, with $\Delta$(inferred property) in absolute units. Points with EW $> 300$, BB $> 0.75$ and cnUV $> 0.3$ are labeled with object IDs (UNCOVER MSA spectroscopic IDs) that are outliers in either inferred parameter space, or have SPS catalog p(z) uncertainties $> 0.1$, which could potentially influence the estimate of Halpha/Balmer Break and cnUV estimates in photometry-only empirical measurements. The highly uncertain p(z) objects are denoted by star symbols.}
\label{fig:delta_2}
\end{figure*}

\begin{figure*}[htb!]
\centering
\includegraphics[page=1,width=1.0\textwidth]{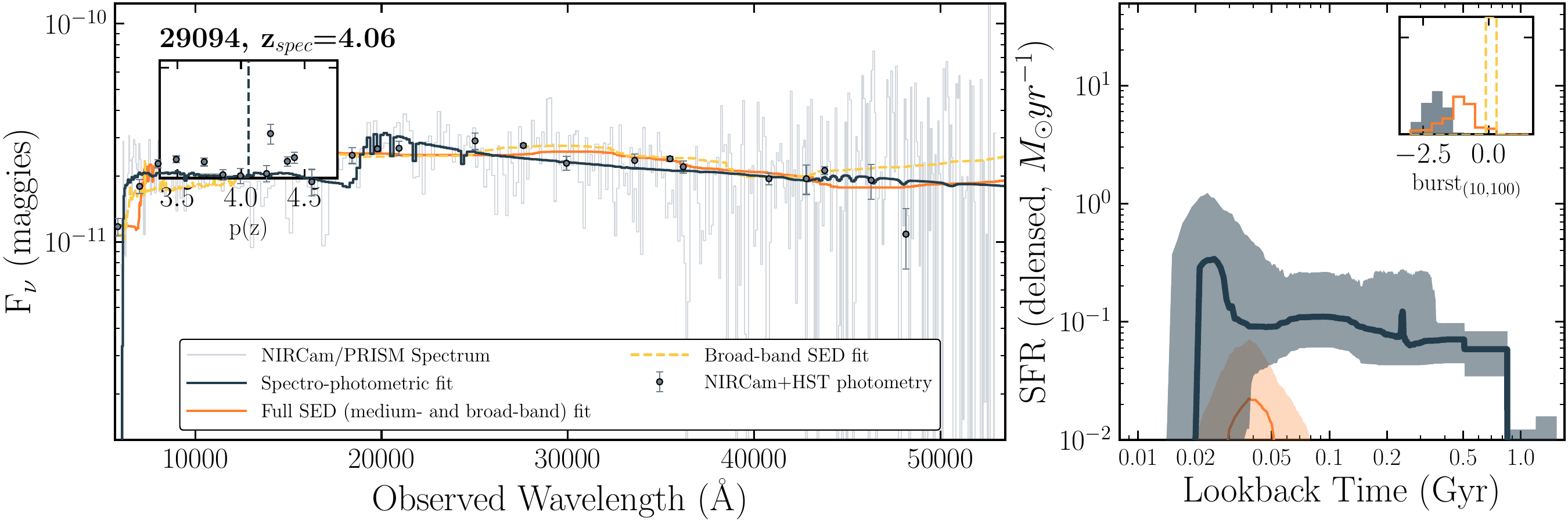}
\includegraphics[page=2,width=1.0\textwidth]{figures/spec_sfh_burst_comparison.pdf}
\includegraphics[page=4,width=1.0\textwidth]{figures/spec_sfh_burst_comparison.pdf}

\caption{\textbf{(Left)} \code{Prospector} SED fits for the UNCOVER napping galaxy sample via \code{Prospector} SPS modeling using the "flexible continuity SFH" model framework. \JWST NIRSpec PRISM spectra (in F$_{\nu}$ maggies units) are shown in grey, and \JWST NIRCam photometry -- broad- and medium-band are plotted in filled grey circles. The maximum a posteriori (MAP) SED models are overlaid, with joint spectroscopy and photometry best fit shown in maroon, photometry-only best fit shown in purple, and broad-band photometry-only fit shown in purple.
\textbf{(Left Inset)} Histograms of posterior distributions of photometric redshifts from SED fitting; the spectroscopic redshift is marked with vertical maroon lines.
\textbf{(Center)} Inferred SFHs of UNCOVER Nappers, as a function of lookback time (Gyr). The SFH measurements from spectrophotometric fits robustly capture the epoch of quenching, and the existence of older stellar populations.
\textbf{(Right)} Posterior distributions of the burstiness parameter log(SFR$_{10}$/SFR$_{100}$) for UNCOVER nappers -- filled maroon histogram is the posterior from spectrophotometric fits, while photometry-only fits posteriors (with and without medium-band photometry) are plotted in hollow histograms). The inferred burstiness of mini-quenched galaxies from spectrophotometric fits are significant outliers in this parameter space; flexible star formation history modeling constrains the burstiness parameter uncertainties reliably, including bimodal solutions. Medium-band photometry-based SED fits may or may not capture quenching, or the same quenching episode as the spectrum-based fits. Note that for 29094, the redshift solution for medium and broad-band fits is incorrect, leading to incorrect SFHs.}
\label{fig:sedfit_2}
\end{figure*}

\begin{figure*}[htb!]
\centering
\includegraphics[page=6,width=0.9\textwidth]{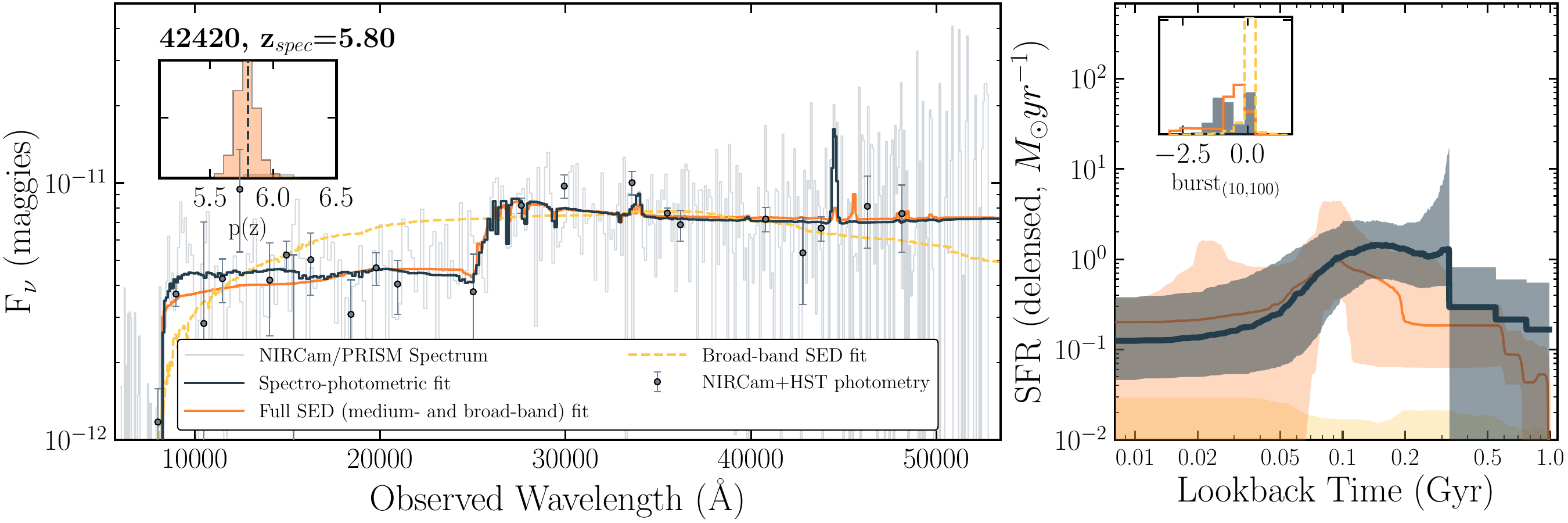}
\includegraphics[page=8,width=0.9\textwidth]{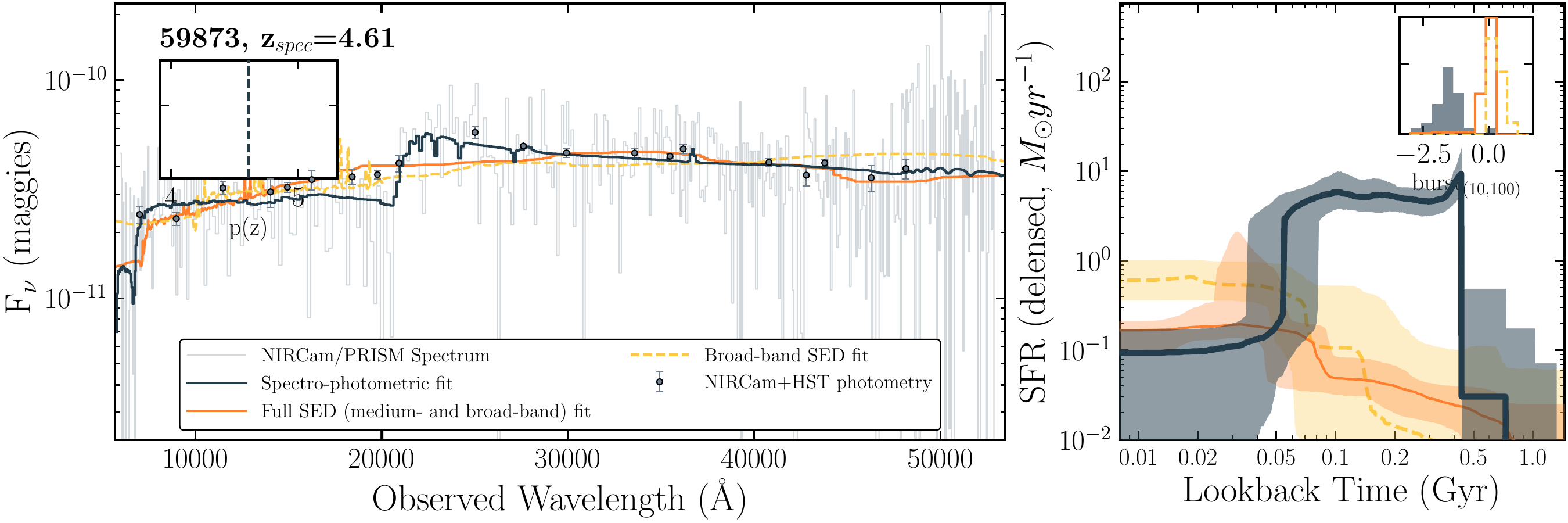}
\caption{Same as Figure \ref{fig:sedfit_2}; continued.}
\label{fig:sedfit_3}
\end{figure*}

See Figures \ref{fig:delta_1} and \ref{fig:delta_2}, where we demonstrate that the mismatch in spectroscopic and photometric measurements of EW(H$\alpha$+[NII]) and cnUV measurements -- $\Delta$(inferred property) -- in all UNCOVER spectra at $z=4-7$ have a weak dependence on the SNR (detection filter --- F277W + F356W + F444W in UNCOVER). See middle column of Figures \ref{fig:delta_1}, where the spectroscopically-confirmed nappers (circles with black outline) with large  $\Delta$(Balmer break) have low SNR measurements for the filter sampling the blue end of the Balmer break (this filter is redshift dependent for each object). In Figure \ref{fig:delta_2}, we mark each object with a large $\Delta$(inferred property) with their MSA ID (and tag large p(z) differences between spec- and phot- measurements with star symbols), further solidifying the evidence for this hypothesis.

The outlier objects here have a mismatch between photometric and spectroscopic measurements due to a host of reasons, e.g., 23608 has an emission line strength mismatch between phot and model (likely due to redshift+filter sampling of OIII or Halpha), while 14411 and 8943 have a mismatch between 16-84 percentile p(z) and DJA v4.4 spec-z ($|z_{spec} - z_{phot}| > 0.1$), and 13416 and 27111 suffer from p(z)-OII-Balmer Break degeneracy in dusty SED solutions. 

\section{Dust correction for the estimation of cnUV and its impact} \label{sec:cnuv_appendix2}

Optical continuum-normalized UV flux, or cnUV, is the new estimator introduced in this work. Here, we use optical continuum as a proxy for a napper's stellar mass (assumed to be mostly dust-attenuation free), while UV flux at 1800$\AA$ -- which correlates with SF at 100-200 Myr timescale -- is attenuated by dust within galaxies. 

Aligning with our philosophy of analysing observations directly, our fiducial cuts use the observed cnUV to remove dusty/maximally quiescent objects. Here, we test the impact of dust-attenuation correction in the UV flux on the measurements of cnUV (and subsequently the selection of nappers). 

We correct our rest-frame 1800\,\AA\ fluxes for dust attenuation using the Small Magellanic Cloud (SMC) extinction curve from \citet{Gordon2003}, which is steeper in the UV than the Milky Way curve and lacks the 2175\,\AA\ bump characteristic of Galactic dust. The SMC curve is particularly suitable for high-redshift star-forming galaxies, which often exhibit similarly steep UV slopes \citep{Calzetti1994, Reddy2018}. We parameterize the extinction curve as $k(\lambda) = R_V \times (-0.11 + 1.21/\lambda_{\mu\mathrm{m}})$ with $R_V = 2.74$ for the SMC Bar \citep{Gordon2003}. 

The UV continuum slope $\beta$ (where $f_\lambda \propto \lambda^\beta$) is calculated by fitting a power law to the rest-frame spectrum between 1268--2580\,\AA, excluding known emission line regions (e.g., Ly$\alpha$ at 1216\,\AA, C\textsc{iv} at 1549\,\AA). For spectroscopically confirmed galaxies, we directly fit the observed continuum using a linear regression in log-log space: $\log(f_\lambda) = \beta \log(\lambda) + C$, where $\beta$ is determined from the slope \citep{Calzetti1994, Meurer1999}. For photometric measurements, we estimate $\beta$ from the flux ratio between tophat filters centered at 1750--1950\,\AA\ and 3750--3950\,\AA\ in the rest frame (akin to other empirical measurements in this work; also see \citealt{mitsuhashi2026}).

We determine the attenuation at 1600\,\AA\ using the \citet{Meurer1999} IRX-$\beta$ relation $A(1600) = 4.43 + 1.99\beta$, which empirically connects UV spectral slope to dust attenuation in local and intermediate-redshift starburst galaxies. This relation has been shown to hold reasonably well for high-redshift star-forming galaxies, though with some scatter depending on SFR/SFHs \citep{Reddy2018, McLure2018, Bouwens2020}. We scale this to our reference wavelength of 1800\,\AA\ using $A(1800) = A(1600) \times [k(1800)/k(1600)]$, where $k(\lambda)$ values are computed from the SMC extinction curve. 

To avoid over-correcting intrinsically blue galaxies with negligible dust content, we implement a floor condition: if $\beta < -2.23$ (corresponding to $A(1600) = 0$ in the \citealt{Meurer1999} relation), we apply no dust correction. This threshold represents the intrinsic UV slope of young, unobscured stellar populations \citep{Calzetti1994}. The dust-corrected flux is computed as $f_{\rm intrinsic} = f_{\rm observed} \times 10^{0.4 A(1800)}$, which is then used to calculate the dust-corrected continuum-normalized UV flux:

\begin{equation}
\mathrm{cnUV_{dust\text{-}corrected}} \equiv \frac{F_{\lambda,\,\mathrm{UV,\,intrinsic}}(1800\,\text{\AA})}{F_{\lambda,\,\mathrm{cont}}}
\end{equation}

where $F_{\lambda,\,\mathrm{cont}}$ is the optical continuum measured in the rest-frame 6230--6950\,\AA\ region, fitted from regions blueward and redward of the H$\alpha$+[N\textsc{ii}] emission lines.

For spectroscopically confirmed UNCOVER nappers at $z=4$--7, this correction results in a median correction factor of 1.2--2$\times$ in the UV flux, with a range of 1.0--3.5$\times$ depending on the measured $\beta$ slope. Galaxies with redder UV slopes ($\beta \sim -1$ to 0) receive larger corrections, while those with intrinsically blue continua ($\beta \sim -2$ to $-2.2$) receive minimal or no correction due to the floor threshold. An empirical cut of $\mathrm{cnUV_{dust\text{-}corrected}} > 0.45$ successfully recovers the same eight spectroscopic nappers identified using the uncorrected threshold of $\mathrm{cnUV} > 0.2$. The updated dust-corrected cuts effectively exclude dusty star-forming galaxies (which have red $\beta$ slopes and require large flux corrections of 2--3$\times$) while applying only minor flux corrections ($\sim0.8-1.2\times$) to nappers with blue UV continua characteristic of recently quenched systems. 

Our photometric napper selection using $\mathrm{cnUV_{dust\text{-}corrected}} > 0.45$, combined with cuts on Balmer break strength ($>1.35$) and H$\alpha$ equivalent width ($<-65$\,\AA), identifies 48 candidate galaxies at $z=4$--7. This sample yields consistent stellar mass functions and fractional number densities within uncertainties compared to the uncorrected sample, demonstrating the robustness of our selection methodology to the choice of dust correction prescription.

\end{document}